# Machine learning on quantum experimental data toward solving quantum many-body problems

Gyungmin Cho and Dohun Kim*

*Department of Physics and Astronomy, and Institute of Applied Physics, Seoul National University, Seoul 08826, Korea*

**Corresponding author: dohunkim@snu.ac.kr*

**Advancements in the implementation of quantum hardware have enabled the acquisition of data that are intractable for emulation with classical computers. The integration of classical machine learning (ML) algorithms with these data holds potential for unveiling obscure patterns. Although this hybrid approach extends the class of efficiently solvable problems compared to using only classical computers, this approach has been realized for solving restricted problems because of the prevalence of noise in current quantum computers. Here, we extend the applicability of the hybrid approach to problems of interest in many-body physics, such as predicting the properties of the ground state of a given Hamiltonian and classifying quantum phases. By performing experiments with various error-reducing procedures on superconducting quantum hardware with 127 qubits, we managed to acquire refined data from the quantum computer. This enabled us to demonstrate the successful implementation of classical ML algorithms for systems with up to 44 qubits. Our results verify the scalability and effectiveness of the classical ML algorithms for processing quantum experimental data.**

Progress in information storage and processing techniques[1–5] has given rise to the generation of large amounts of data, and the use of machine learning (ML) to process these data is being actively explored in biology[6], chemistry[7], and physics[8]. Areas of potential application of ML in physics include the study of many-body physics. One of the interesting problems is the prediction of the ground-state properties of a given Hamiltonian, for example the electronic structure Hamiltonian. The other intriguing problem entails the exploration of the boundaries between different quantum phases, which may enable the identification of exotic quantum phases such as high-temperature superconductivity. Widely used classical algorithms[9,10], however, in spite of their many successful applications, have fundamental limitations in approximating strongly interacting systems.

Although quantum computers are expected to excel at solving quantum chemistry and many-body physics problems, current devices are still prone to errors, which compromise the accuracy of results. While quantum error correction is believed as a solution[11,12], the large-scale operation is not immediately realizable[13]. Hybrid approaches in which classical computers are combined with quantum computers as to circumvent this issue have been introduced, but it requires the conversion of quantum data into classical data through the measurement process[14].

Research aimed at broadening the utility of the hybrid approach leads to quantum state learning methods such as classical shadow[15–17]. The ability to convert quantum states into classical forms naturally allows for their use as data in classical ML, and recent research has focused on theoretical analyses of the performance in many-body physics applications[18–21]. One advantage in this direction is that it allows well-developed classical devices and ML techniques to be leveraged. Additionally, under widely believed complexity conjectures, the combined use of data from quantum computers and learning on classical computers is strictly more powerful than solving problems using classical computers only[18,22]. Despite the aforementioned advantages, prevailing errors in data from quantum computers limit the range of problems that are addressable and challenge the scalability as the system size increases[23–25]. Thus, appropriate quantum error mitigations (QEMs) are necessary during and after data acquisition.

Here, we experimentally verify the applicability of the hybrid approach to problems of interest in many-body physics. Using data from quantum computers, we implement classical ML algorithms that have been theoretically studied[18] to solve problems related to the prediction of ground state properties and the classification of quantum phases, as illustrated in Fig. 1. These problems can be considered as regression and classification in traditional ML, respectively. For regression, because the prediction of accurate values is required, various QEMs were used, which resulted in the application of accurate ML models to a 12-qubit system. Regarding the classification task, we expanded on the previous work[26,27] of distinguishing the Symmetry Protected Topological (SPT) phase[28] by performing the classification task in a more general setting and increasing the system size up to 44 qubits. Also, with the help of a measurement-assisted state preparation method[29], which enabled the generation of suitable training data, we demonstrated successful phase classification between topologically ordered and trivial phases of a system comprising as many as 25 qubits, thereby confirming the scalability and applicability of the ML algorithms. We conducted our experiments on a device consisting of fixed-frequency superconducting transmon qubits provided by IBM Cloud. Detailed error statistics on the hardware are presented in the Supplementary Information Note 1.

## Classical shadow

Despite many efforts to reduce the sample complexity for quantum state tomography (QST), exponential scaling of the sample complexity in terms of the system size ($n$) is unavoidable[30].

The central idea of classical shadow[15] is the following: by focusing on the expectation value of few-body operators, rather than on entire information of the state, we could circumvent the exponential scaling of the sample complexity. In addition to linear functions $f_{\text{linear}}(\rho) = \text{Tr}(O\rho)$, nonlinear functions such as $f_{\text{nonlinear}}(\rho) = \text{Tr}(O\rho^{\otimes k})$ can also be estimated. Utilization of the classical shadow as the data for ML would be expected to enable the ML model to learn the nonlinear properties of the state[18,22].

We obtained the classical shadows of the state by applying a unitary transformation sampled from a random unitary ensemble, followed by the measurement. By repeating this process $T$ times, we can obtain $S_T(\rho) = \{(b^{(t)}, U^{(t)})\}_{t=1}^{T}$ as the experimental results, where $b^{(t)} \in \{0, 1\}^n$ is the measurement outcome, $U^{(t)} = \otimes_{i=1}^{n} U_i^{(t)}$ where each $U_i^{(t)}$ is sampled from the Haar measure over the unitary group $\mathbb{U}(2)$. As a result, the unbiased estimator $\hat{\sigma}_T(\rho)$ can be written as

$$\hat{\sigma}_T(\rho) = 1/T \sum_{t=1}^{T} \{\otimes_{i=1}^{n} (3U_i^{(t)\dagger} |b_i^{(t)}\rangle\langle b_i^{(t)}| U_i^{(t)} - I_2)\}$$

where $\mathbb{E}_{U,b}(\hat{\sigma}_T(\rho)) = \rho$ and $I_2$ is the $2 \times 2$ identity matrix (1).

In practice, classical shadow estimations sometimes lead to inferior results compared to direct measurement due to the restriction on available QEMs resulting from the randomized measurements (RM)[16] for the classical shadow. Here, direct measurement was used for the regression, whereas classical shadow representation was employed for the classification in which errors were tolerable to a certain extent, but nonlinear properties were necessary.

**Case 1: Predicting the properties of the ground state**

In our first experiment, we aimed to predict the properties of the ground state, specifically by focusing on learning that $f(x) = \text{Tr}(O\rho(x))$ ($\rho(x)$ is the ground state of the parameterized Hamiltonian $H(x)$). By mapping an input vector $x$ to a high-dimensional space through a feature map $\varphi: x \in R^m \rightarrow R^{m_\varphi}$, functions $f(x)$ can be approximated by $w^T\varphi(x)$ where $w$ is a model parameter. However, if the dimension of the feature space ($m_\varphi$) is too large, it is impractical to conduct calculations directly using feature vector $\varphi(x)$. Instead of having to process high-dimensional vectors, a relatively simple relation known as the kernel trick[31] $k(x, x') = \varphi(x)^T\varphi(x')$ can be used. Among the many available algorithms, we used kernel ridge regression (KRR) and a closed-form expression for predicting $f(x_{new})$ based on $N_{data}$ samples given by

$$\hat{f}(x_{new}) = \sum_{i=1}^{N_{data}} \sum_{j=1}^{N_{data}} k(x_{new}, x_i)(K + \lambda I)_{ij}^{-1} f(x_j) \tag{2}$$

where $\lambda$ is the hyperparameter, $K_{ij}(= k(x_i, x_j))$ is a kernel matrix, and $I$ is the $N_{data} \times N_{data}$ identity matrix.

We selected the 1D nearest-neighbor (NN) random hopping system $H_{hop}(x)$ with 12 sites ($n$ = 12) (illustrated in Fig. 2a) as a benchmark for the learning task. The number of parameters of the system increases linearly as $n$ grows. Although it may be challenging to train the system with many parameters, after training, the ML model would have the extended interpolation regions for inferences. The ground state of $H_{hop}(x)$ was prepared on a quantum computer by using the Givens rotations[32,33] in Fig. 2b. To reduce errors, apart from diverse error mitigation methods (Dynamical Decoupling, Pauli twirling, McWeeny purification)[34,35], we implemented a parity measurement by recompiling the circuit. Further information is provided in the Supplementary Information Note 3.

We obtained the expectation values corresponding to the site correlations $\langle a_i^\dagger a_j \rangle$ (= 1/4($X_i$-i$Y_i$)$Z_{i+1}$..$Z_{j-1}$($X_j$+i$Y_j$) by employing a Jordan-Wigner transformation, where $X_i$, $Y_i$, and $Z_i$ are the Pauli operators at site $i$), and constructed a correlation matrix of which each ($i$, $j$)-element corresponds to $\langle a_i^\dagger a_j \rangle$. We uniformly sampled the hopping rate $x \in [0, 2]^{n-1}$ and performed 20,000 measurements to obtain each $\langle a_i^\dagger a_j \rangle$. Then, we used ML with the aim of predicting the correlation matrix for a new ground state at $x_{\text{new}}$. We collected 200 data points from the quantum computer to train the ML model. The performance of the trained model was evaluated using 10,000 test data obtained by Exact Diagonalization (ED). The correlation matrix predicted by the ML model has reasonable similarity to the exact values (Fig. 2c). As shown in Fig. 2d, the model achieved an average root-mean-square error (RMSE) of 0.0168 on the test data, and the decrease in the RMSE as the number of training data points increases confirms the importance of the data in ML for predicting the ground state properties. The red dotted line represents the RMSE of the error-mitigated training data, which becomes 0.0244 in the case of raw data. We observed the scaling relationship $\log(N_{\text{data}}) \sim O(1/\text{RMSE})$ of the trained model (inset in Fig. 2d), even in the multi-phase case, which is guaranteed as an optimal scaling in the single-phase case[18].

The trained ML model was subsequently used to predict the ground state properties of the Su–Schrieffer–Heeger (SSH) Hamiltonian (Fig. 2a), and as seen in Fig. 2e-g, the results confirmed the ability of the ML model to predict not only the correlation matrix but also edge correlations originating from the topological properties. We therefore succeeded in experimentally confirming one of the promising applications of the ML approach, namely that the properties of other systems of interest can be estimated without performing additional experiments. Details of the noisy circuit simulations can be found in the Supplementary Information Note 5.

**Case 2: Classifying quantum phases**

In the second experiment, we classified quantum phases in many-body physics by using principal component analysis (PCA) and a support vector machine (SVM) as classical ML algorithms[31]. To generate data for ML, we prepared the fixed-point state of a given phase on the quantum computer and applied a local random unitary to generate different states within the same phase as the training data[36]. The advantage of this approach is that it allows for model-independent data acquisition, thereby reducing the biases in the training data[37]. With a classical shadow that contains sufficient information to compute the nonlinear properties of the state, we employed a shadow kernel[18] defined by

$$k_{\text{shadow}}(S_T(\rho),\ S_T(\tilde{\rho})) = \exp\{\tau/T^2 \textstyle\sum_{t,t'}^{T} \exp[\gamma/n \sum_{i=1}^{n} Tr(\sigma_i^{(t)}\tilde{\sigma}_i^{(t')})]\} \quad (3)$$

where $\sigma_i^{(t)} = 3U_i^{(t)\dagger}\left|b_i^{(t)}\right\rangle\left\langle b_i^{(t)}\right|U_i^{(t)} - I_2$ and $\tau, \gamma > 0$ are hyperparameters. Each $k_{\text{shadow}}$ can be efficiently evaluated using $O(nT^2)$ computation time. Further explanations on the shadow kernel can be found in the Supplementary Information Note 2.

**Case 2-1:** Distinguishing a short-range entangled state from a trivial one

Without symmetry, both the SPT phase, having short-range entanglement, and the trivial phase exhibit trivial order, and the states from these two phases can be connected via constant-depth local unitary (LU) transformations[36] (Fig. 3a). However, in cases in which the applied unitary preserves a specific symmetry, the constant-depth circuit protecting the symmetry is known to be non-existent[38]. As a fixed-point state in the SPT and trivial phases, we utilized the ground state of $H_{ZXZ} = -\sum_i Z_{i-1}X_iZ_{i+1}$ and $H_X = -\sum_i X_i$ with a 44-site periodic boundary condition, respectively, and examined whether the ML model can distinguish the SPT phase protected by $\mathbb{Z}_2 \otimes \mathbb{Z}_2$ symmetry generated by $X_{\text{even(odd)}} = \prod_{i=\text{even(odd)}} X_i$ or time-reversal symmetry (TRS) $\mathcal{T} = (\prod_i X_i)K$, where $\mathbb{Z}_2$ is a second-order group and $K$ denotes complex conjugation. Specifically, we do not assume that the system is translationally invariant when applying a symmetric random unitary[37] (Fig. 3b). This means that attempts to measure the string order parameters (SOP) to distinguish the SPT phase are highly likely to fail, as shown in Fig. 3c[27,28]. However, ML using classical shadows to generate data could increase the probability of classifying the SPT phase because it contains nonlinear information about the state[39].

We obtained 20 data points for each phase in the form of classical shadows using $T$ = 100, and used half of these points as training data to identify the phase boundary and the other half as test data. For classical ML, PCA with a shadow kernel was used to reduce the dimensionality of the data, and the phase boundary was obtained using SVM with a Gaussian kernel. Fig. 3d shows that the ML model can distinguish between the SPT phase protected by $\mathbb{Z}_2 \otimes \mathbb{Z}_2$ symmetry or TRS with high accuracy. However, if the symmetry is not respected, the two phases can be connected by a constant depth local unitary, which would complicate classification. Fig. 3e and 3f show the distribution of the test data and phase boundary of the trained ML model for the $\mathbb{Z}_2 \otimes \mathbb{Z}_2$ symmetry and TRS. In the case of TRS, the symmetric local unitary gates are less restrictive than in the $\mathbb{Z}_2 \otimes \mathbb{Z}_2$ symmetry, which permits the generation of a wider variety of states from the fixed-point state to ultimately give rise to a more dispersed data distribution.

Lastly, we attempted to distinguish the quantum phases of the Cluster-Ising Hamiltonian $H_{\mathrm{CI}} = -J\sum_i Z_i X_{i+1} Z_{i+2} - h_1 \sum_i X_i - h_2 \sum_i X_i X_{i+1}$, for which multiple quantum phases exist depending on the values of $h_1$ and $h_2$ (with $J$ = 1). [26,40,41] Utilizing the same experimental data under $\mathbb{Z}_2 \otimes \mathbb{Z}_2$ symmetry, we trained a ML model and attempted to distinguish between the SPT and trivial phases among a total of 40 test data points from $H_{\mathrm{CI}}$—20 from each phase—obtained by density matrix renormalization group (DMRG) simulation[42,43], as shown in Fig. 3g. These results showed that all test data points, except for one, were correctly classified into appropriate phase, demonstrating the applicability of ML with the classical shadow.

**Case 2-2:** Distinguishing a long-range entangled state from a trivial one

Topologically ordered states, having long-range entanglement, directly prepared on quantum devices have been confirmed by measuring the non-zero expectation value of topological string operators[44] or topological entanglement entropy (TEE)[45–47]. Here, we classified topologically ordered phases using the ML model trained by data. We utilized a surface code (a planar version of the toric code), which is well known to have a $\mathbb{Z}_2$ topological order, as a fixed-point state for a topologically ordered phase and a random product state for a trivial phase. The method that was previously used for preparing the topologically ordered state[47] requires hardware-specific qubit connectivity conditions, which would otherwise incur additional swap gates and lower the state preparation fidelity. In addition, the required circuit depth for these methods increases as $O(d_{code})$ where $d_{code}$ is the code distance of the surface code[47]. We avoided these problems by utilizing the measurement-assisted state preparation[29] method instead. Specifically, after applying a sequence of unitary transformations in Fig. 4a, we performed a projection by measuring some ancilla qubits and applied Pauli $Z$ or an identity gate adaptively to certain data qubits conditioned on the measurement results. Despite the non-deterministic nature of the projection, it is always possible to prepare $|0_L\rangle$ of surface code as follows at O(1) circuit depths.

$$|0_L\rangle = 1/\sqrt{2^{N_p}} \prod_p (I + B_P)|0\rangle^{\otimes n} \quad (4)$$

Here, $B_P = \prod_{i \in p} X_i$ is a $X$-plaquette operator where each $X$-plaquette is shaded in blue in Fig. 4a, $N_p$ is the number of $X$-plaquettes. In addition, we applied the adaptive virtual gate to classical shadows to eliminate possible errors arising from idle data qubits during the measurement of the ancilla qubits. This flexibility is another advantage of using the classical shadow as data for classical ML. Further details on this method can be found in the Supplementary Information Note 3.

We prepared a $d_{code} = 5$ surface code and applied tensor products of the random single qubit gates to generate data for the topologically ordered phase. For the trivial phase, we prepared a random product state and applied local random unitary at depth ($d_{LU}$) from 0 to 5 (Fig. 4b). We obtained 20 data points from each phase using $T = 300$. The ML model was trained by following the same procedure as in Case 2-1. As shown in Fig. 4c, it was possible to distinguish the phases by unsupervised learning when $d_{\mathrm{LU}} = 0$. The distribution of the trivial phases spread out as the $d_{\mathrm{LU}}$ increased, and the phase boundary faded beyond $d_{\mathrm{LU}} \sim (d_{\mathrm{code}} - 1)/2 = 2$. To verify that the observed phenomenon originated from a robust topological property, we carried out the same experiments for a randomly sampled state with the same gate complexity for preparing $|0_{\mathrm{L}}\rangle$ but without any topological order. The results indicated that a phase boundary does not exist, even when $d_{\mathrm{LU}}$ is small (Fig. 4d).

In comparison to previously studied direct measurements of the topological order parameter, measuring topological string operators[44] within error $\varepsilon$ only required $O(1/\varepsilon^2)$ measurements; however, deviation of the state from the fixed-point state degraded its classification ability[18]. This could be avoided by evaluating the nonlinear properties such as TEE[45,46] but it would require more measurements than were made in our case.

**Discussion:** Unlocking the black-box of the ML model

A drawback of ML-based analyses is the black-box nature of the ML model, which makes it challenging to understand which aspects of the data are utilized in these analyses. To address this issue, we conducted experiments to extract relations among the data that were used by the ML model. In experiment, we collected 10 data points each from both the topologically ordered and trivial phases consisting of 9 qubits with local random unitary applied to a fixed-point state. Then, we used SVM on the feature space formed by the feature map $\varphi(\rho)$, which transformed the input quantum state into a vector consisting of the subsystem Renyi-2 entanglement entropy (Fig. 5a). ML in this case corresponds to finding an optimal linear combination of subsystem Renyi-2 entanglement entropy. Details of the mapping procedure and a specific form of the ML classifier can be found in the Supplementary Information Note 5. The evaluation of the trained ML model presented in Fig. 5b was conducted using 100 datasets (each set consisted of 6,000 states, generated in the same way as the training data, but through classical simulation). We observed that the phase classifier derived from the ML model distinguishes quantum phases more effectively than the previously used nonlinear order parameter of Renyi-2 TEE[47]. This result is attributed to the intentional introduction of errors in training data, a technique named as data augmentation[48]. Furthermore, a comparison of the prediction error of the model with and without measurement error mitigation (MEM) in the training data enabled us to confirm that appropriate error mitigation techniques can help improve the ML performance.

**Conclusion/Outlook**

Our results highlight an interesting aspect, namely the use of classical ML to process quantum experimental data with problem-specific error-reducing procedures for studying quantum many-body physics. Extensions of our work and interesting future directions can be outlined as follows. Rather than conducting the measurements immediately after preparing the quantum state, compressive quantum transformations preserving an important feature of the state before measurements could lower the dimensionality of the quantum state, which could reduce the computational time for classical ML. Additionally, investigating the non-equilibrium properties using ML based on data obtained from dynamical simulations of quantum systems[49] would be a promising generalization of our work. Consequently, we anticipate that future work based on our results would continue to extend the useful applications of quantum devices before fault-tolerant quantum computers.

## Methods

### Kernel Ridge Regression

To predict the properties of the ground state from the measured expectation values, we utilized the classical ML algorithm, Kernel Ridge Regression (KRR). KRR aims to estimate the value of a real-valued function $f(x)$ for a new input vector $x_{\mathrm{new}}$ using the given data $\{(x_i, f(x_i))\}_i$. Here, $f(x_{\mathrm{new}})$ was approximated as $f_{\mathrm{ML}}(x_{\mathrm{new}}; w) = w^{\mathrm{T}}\varphi(x_{\mathrm{new}})$ by employing $w$, which minimizes the $L_2$ loss with the regularization $C(w) = \frac{1}{2}\sum_{i=1}^{N_{data}} \left| w^T\varphi(x_i) - f(x_i) \right|^2 + \frac{1}{2}\|w\|_2^2$. As the feature vector $\varphi(x)$ typically resides in high-dimensional space, we used the kernel trick instead of directly calculating the high dimensional feature vector $\varphi(x)$.

### Kernel Principal Component Analysis

For reducing the dimensionality of data, we applied kernel principal component analysis (PCA) to quantum states represented as classical shadows, using the shadow kernel. PCA was implemented by defining the covariance matrix $C = \frac{1}{N_{\mathrm{data}}}\sum_{i=1}^{N_{\mathrm{data}}} \varphi(x_i)\varphi(x_i)^{\mathrm{T}}$ and by diagonalizing $C$, which led to principal axes that could describe the data with the minimum possible number of variables. But, mapping an input vector $x$ into a large dimensional vector through the feature map $\varphi : x \in R^m \rightarrow R^{m_\varphi}$, as in our case, would cause the direct diagonalization of $C$ to become computationally challenging. Using the kernel trick as a workaround, by diagonalizing the kernel matrix $K$, defined by $K_{ij} = k(x_i, x_j) = \varphi(x_i)^{\mathrm{T}}\varphi(x_j)$ for indices $i, j$ ranging from 1 to $N_{\mathrm{data}}$, we can derive the expression for principal axes instead of diagonalizing covariance matrix $C$. Using kernel PCA, we effectively reduced the dimensionality of the data, which led to the clustering of data points within the same quantum phase.

**Error-reducing procedures**

We utilized various methods to reduce the errors arising from the data acquisition process in quantum computing. Both Dynamical Decoupling (DD) and Pauli twirling (PT) were employed in all experiments. Additional techniques such as particle number conservation, McWeeny purification, and parity measurement by recompiling the circuit were used to train ML model for the 1D nearest-neighbor random hopping system. In the task of distinguishing quantum phases, we were able to eliminate the swap gate overhead induced by the hardware qubit connectivity through the measurement-assisted state preparation method and reduce the errors from idle data qubits by using the virtual (adaptive) gate on the classical shadow. Furthermore, in the experiment that involved extracting the ML classifier from the trained ML model, we utilized measurement error mitigation, which improved the performance of the trained ML model. Detailed explanations for each method can be found in the Supplementary Information Note 3.

**Classical simulation for ground states**

The test data in Fig. 3g were obtained by using matrix product state (MPS) representation and density matrix renormalization group (DMRG) to compute the ground state of $H_{\mathrm{CI}} = -J\sum_i Z_i X_{i+1} Z_{i+2} - h_1 \sum_i X_i - h_2 \sum_i X_i X_{i+1}$. In our simulation, we utilized a bond dimension of $\mathcal{X} = 100$ for a system of 44 qubits and employed a perfect sampling method to enable the simulation of randomized measurements to obtain the classical shadows of the prepared quantum state represented by MPS.

**Hardware characteristics**

We used the quantum hardware (*ibm_sherbrooke*) with 127 qubits available through IBM Cloud. The backend is composed of fixed-frequency transmon qubits where each qubit, embedded in a heavy-hexagonal lattice, is directly connected with two or three other qubits. Detailed device characteristics are summarized in the Supplementary Information Note 1.

**Data availability**

The data that support the findings of this study are available from the corresponding author upon request.

**Acknowledgments**

This work was supported by a National Research Foundation of Korea (NRF) grant funded by the Korean Government (MSIT) (No. 2019M3E4A1080144, No. 2019M3E4A1080145, No. 2019R1A5A1027055, RS-2023-00283291, SRC Center for Quantum Coherence in Condensed Matter RS-2023-00207732, quantum computing technology development program No. 2020M3H3A1110365, and No. 2023R1A2C2005809), a Korea Basic Science Institute (National Research Facilities and Equipment Center) grant funded by the Ministry of Education (No. 2021R1A6C101B418). Correspondence and requests for materials should be addressed to DK (dohunkim@snu.ac.kr ).

**Author contributions**

GC conceived the project, performed cloud-based experiments, conducted classical error simulations, and analyzed the data. GC prepared the manuscript. DK supervised the project.

**Competing interests**

The authors declare no competing interests.

# Figures

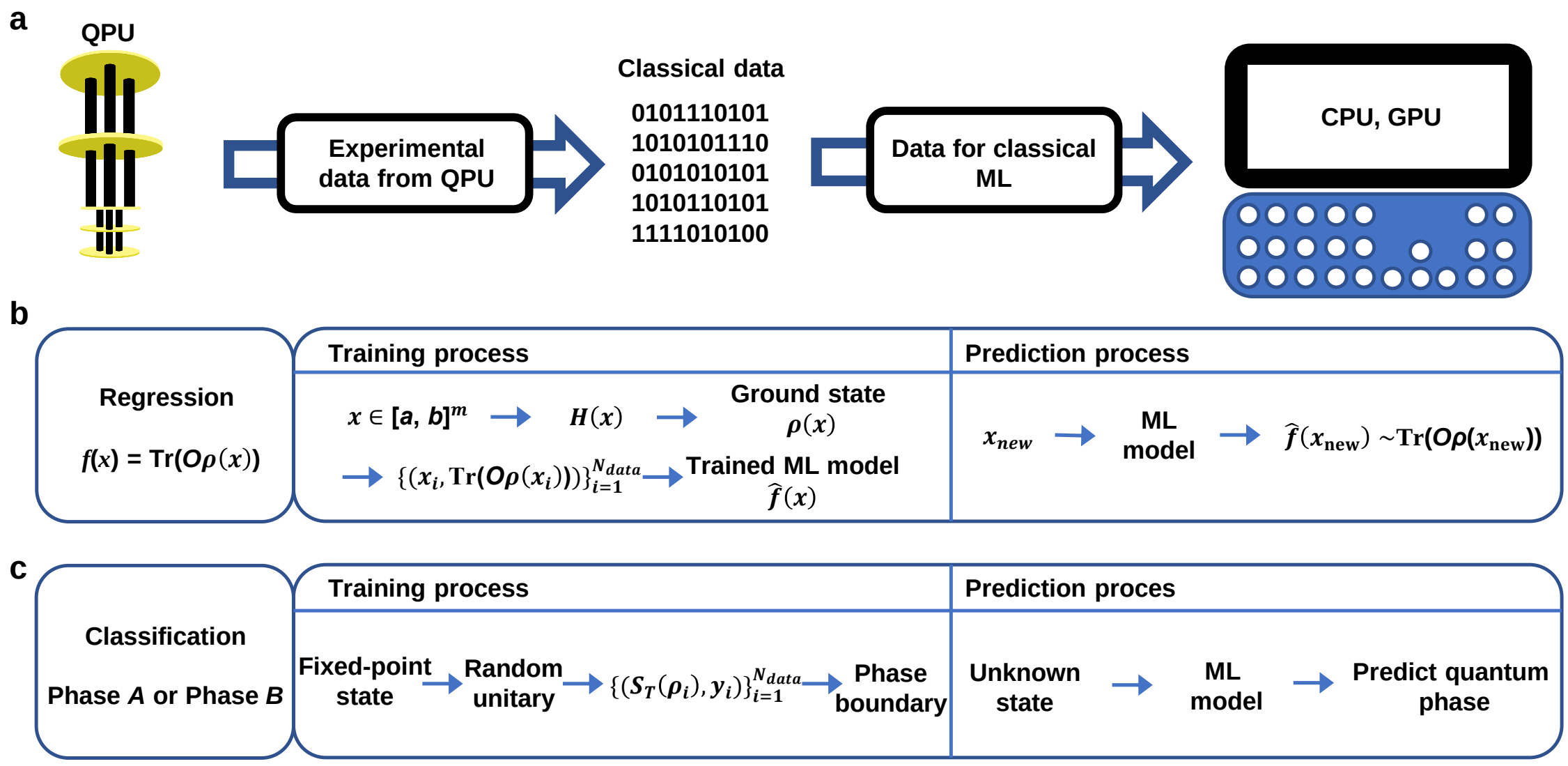

**Fig. 1 | Classical machine learning (ML) on quantum experimental data. a**, Conversion of the information represented in a quantum state into a classical form for the application of classical ML. The training data were the expectation values measured via the quantum computer or the classical shadows of the quantum state. **b**, Predicting the properties of the ground state of a given Hamiltonian casts to regression in classical ML. The objective of the regression is to approximate the target function $f(x) = \mathrm{Tr}(O\rho(x))$ as closely as possible by trained ML model using training data $\{(x_i, \mathrm{Tr}(O\rho(x_i))\}_{i=1}^{N_{\mathrm{data}}}$. After training, when a new input $x_{\mathrm{new}}$ is given, the trained ML model outputs $\hat{f}(x_{\mathrm{new}})$ as a guess for $f(x_{\mathrm{new}})$. **c**, In classification, the objective is to distinguish a particular quantum phase from others by training a ML model that gives rise to the phase boundary on a relevant space. After training, when a new state is given, the trained ML model assigns a phase of the state by considering its position relative to the phase boundary on the space.

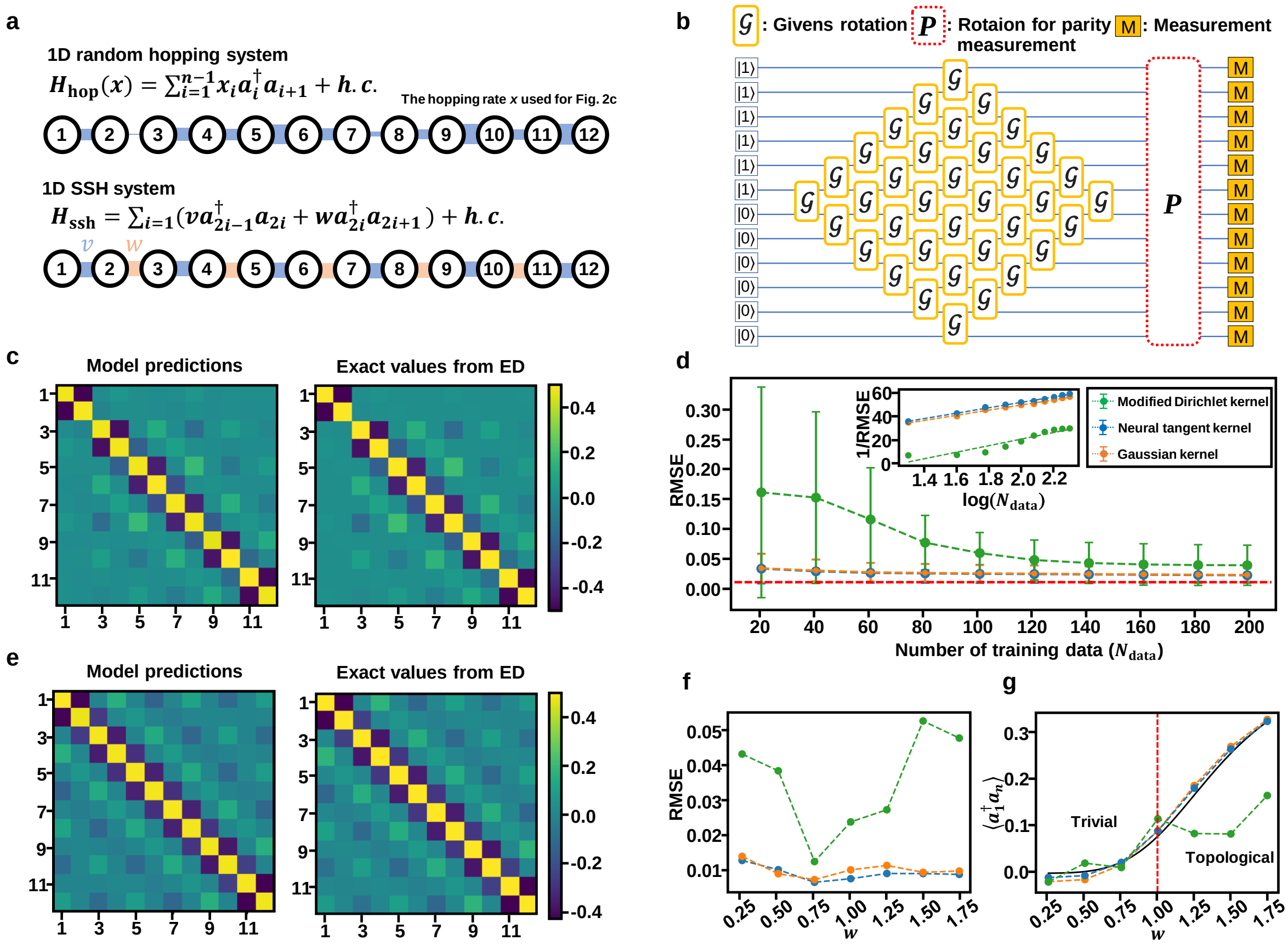

**Fig. 2 | Predicting ground state properties. a**, Hamiltonian for random hopping system and Su–Schrieffer–Heeger (SSH) system. $x_i$ in $H_{\mathrm{hop}}$ is sampled uniformly from [0, 2]. **b**, Ground state preparation circuit. Gates labeled $\mathcal{G}$ and $P$ represent a Givens rotation and a basis rotation for parity measurement, respectively. **c**, Correlation matrix for random hopping model. Results from ML prediction using the Gaussian kernel and exact values from exact diagonalization (ED) are shown on the left and right, respectively. **d**, Prediction error. Model performance is measured by the root-mean-square error (RMSE). The error bars represent the standard deviation of the RMSE over the 10,000 test data points. The dotted red line means the average RMSE for the training data. Inset shows the relationship between the log($N_{\mathrm{data}}$) and 1/RMSE. **e**, Correlation matrix for 1D SSH system ($v$ = 1, $w$ = 1). Results from ML predictions using Gaussian kernel and exact values from ED are shown on the left and right, respectively. **f**, Prediction error. RMSE from different kernels while keeping $v = 1$ and changing $w$ in intervals

of 0.25 from 0.25 to 1.75. **g**, Edge correlation. The edge correlation $\langle a_1^{\dagger} a_n \rangle$ is measured between sites located at either end of a chain. The dotted red line is the known phase boundary between the trivial and topological phases. The black curve indicates exact values obtained by ED. In (**d**), (**f**), (**g**), the green, blue, and orange points each indicate the results predicted by the ML model trained by a (modified) Dirichlet kernel, neural tangent kernel, and Gaussian kernel, respectively.

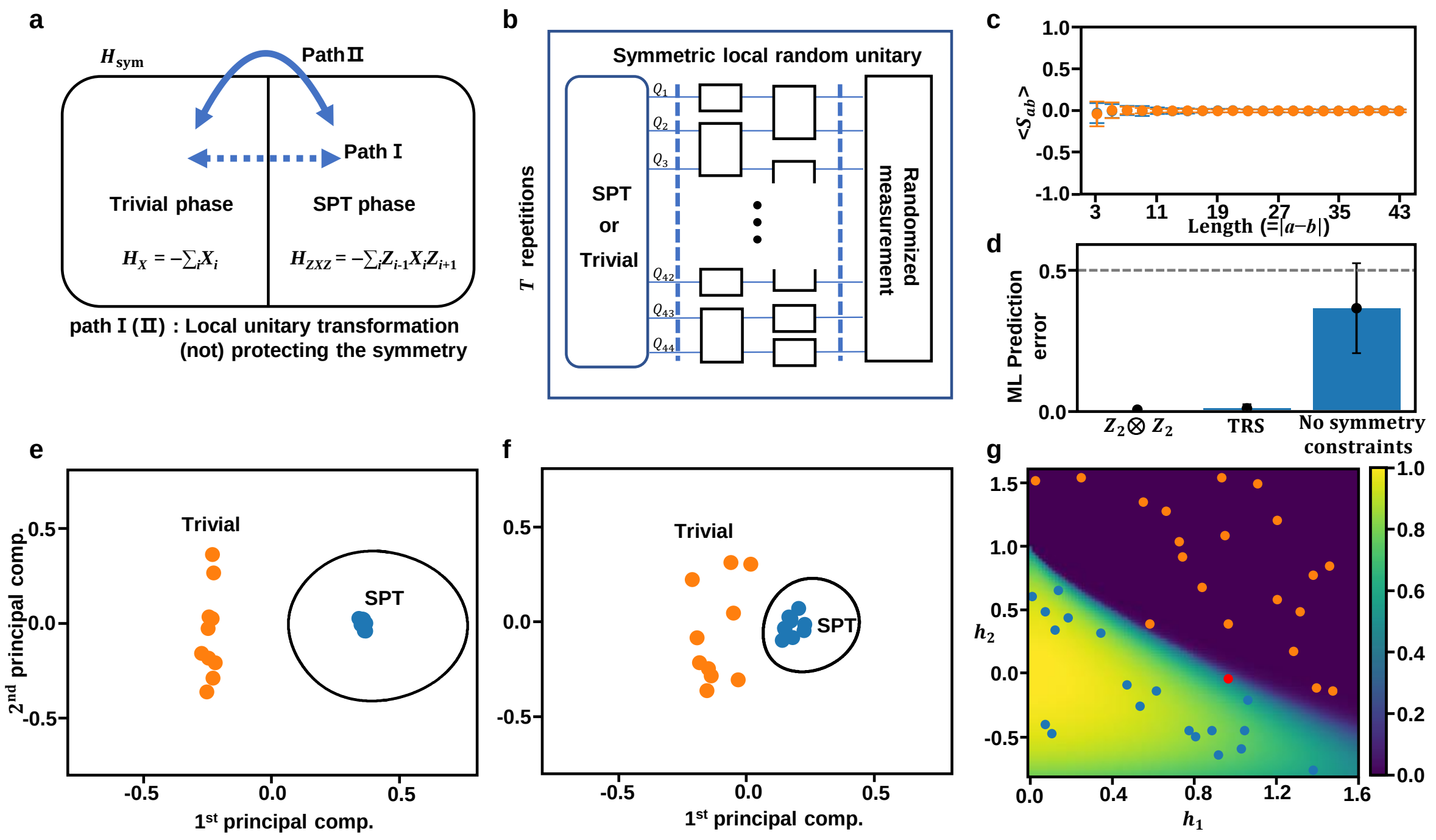

**Fig. 3 | Experimental results for distinguishing between symmetry protected topological (SPT) and trivial phases.** **a**, Trivial and SPT phases. When considering a set of gapped symmetric Hamiltonians only, the two phases cannot be connected by a constant depth symmetric local unitary (Path I). However, without consideration of the symmetry, there exists a constant depth local unitary circuit (Path II) connecting them. **b**, Generation of data. The process involves applying two layers of symmetric local random unitary to a fixed-point state associated with a particular phase, followed by randomized measurement for classical shadows. Random gates used can be found in the Supplementary Information Note 5. **c**, String order parameter (SOP) ($S_{ab} = Z_a X_{a+1} X_{a+3} \dots X_{b-3} X_{b-1} Z_b$) of various lengths measured by direct measurements on the quantum computer. The blue and orange dots correspond to the SPT and trivial phases, respectively. The error bars represent the standard deviation of 300 SOPs. **d**, Classification errors. The chart shows the classification error of the ML model for cases with $\mathbb{Z}_2 \otimes \mathbb{Z}_2$, TRS, and without symmetry constraints. The error bars represent the standard deviation from 10 instances. **e, f**, 2D representation of the test data. Each figure corresponds to $\mathbb{Z}_2 \otimes \mathbb{Z}_2$ (**e**) and time-reversal symmetry (TRS) (**f**). **g,** Classification results of $H_{\mathrm{CI}}$. The

background represents the expectation values of a SOP measured from the ground states of the $H_{\mathrm{CI}}$, and these values are used to determine the ground truth of the quantum phase. For the 40 randomly sampled test data (marked by circles), each phase predicted by the trained ML model is colored in blue for the SPT phase and orange for the trivial phase. The red dot represents a point where the model incorrectly predicted the phase.

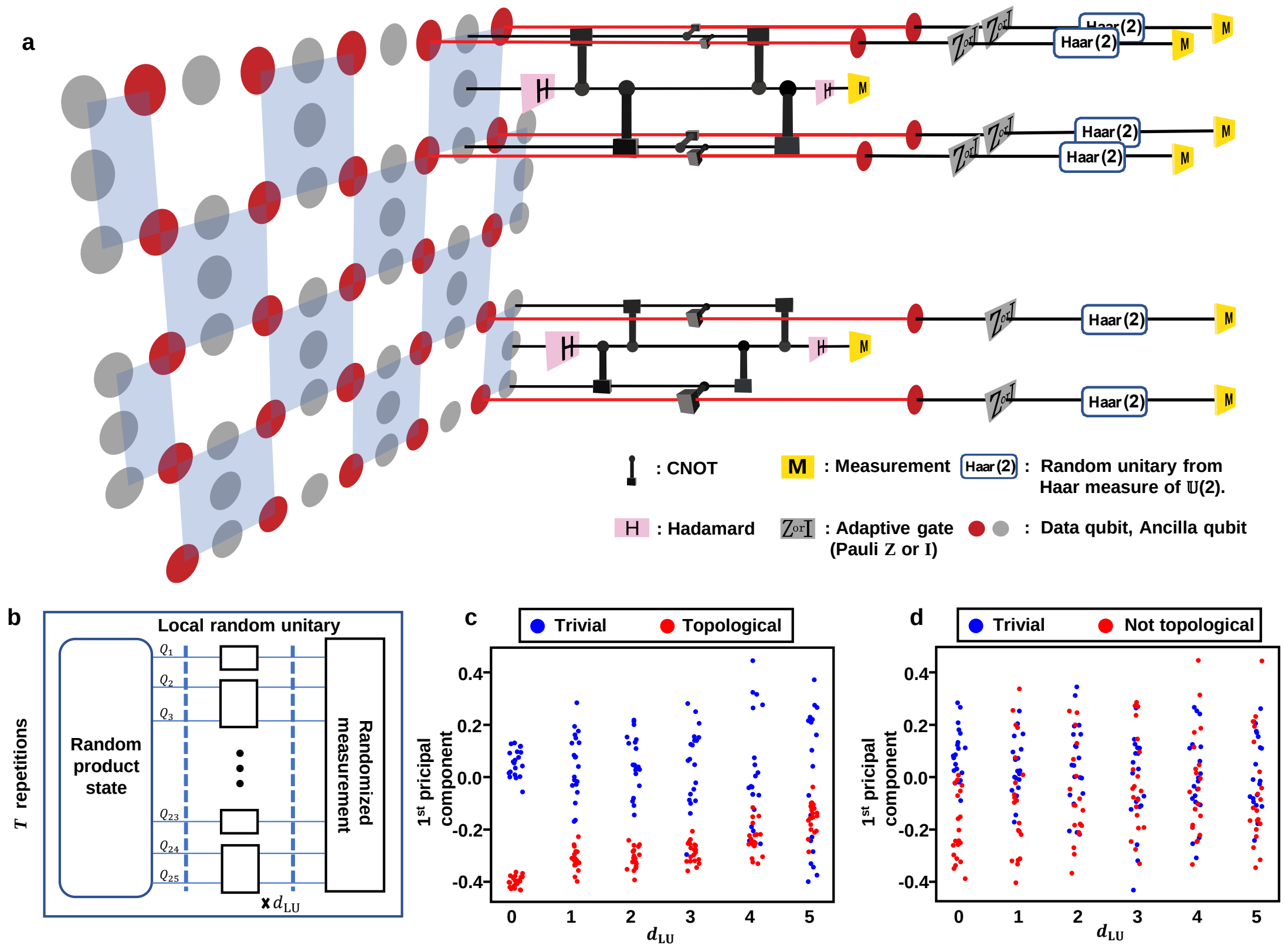

**Fig. 4 | Experimental results for distinguishing between topologically ordered and trivial phases. a**, Schematic diagram for measurement-assisted state preparation method on the heavy-hexagonal lattice. The application of a sequence of unitary transformations and adaptive gates depending on the measurement results of some ancilla qubits was followed by randomized measurements for classical shadows. The blue shaded areas represent the *X*-plaquettes. **b**, Generation of data in the trivial phase. States in this phase are generated by applying local random unitary to the product state at varying circuit depth ($d_{\mathrm{LU}}$). Detailed information about the local random unitary that was used is provided in the Supplementary Information Note 5. **c**, 1D projection of data. Using kernel principal component analysis with the shadow kernel in (3), each data point was projected onto the first principal axis. **d**, 1D projection of data without topological order.

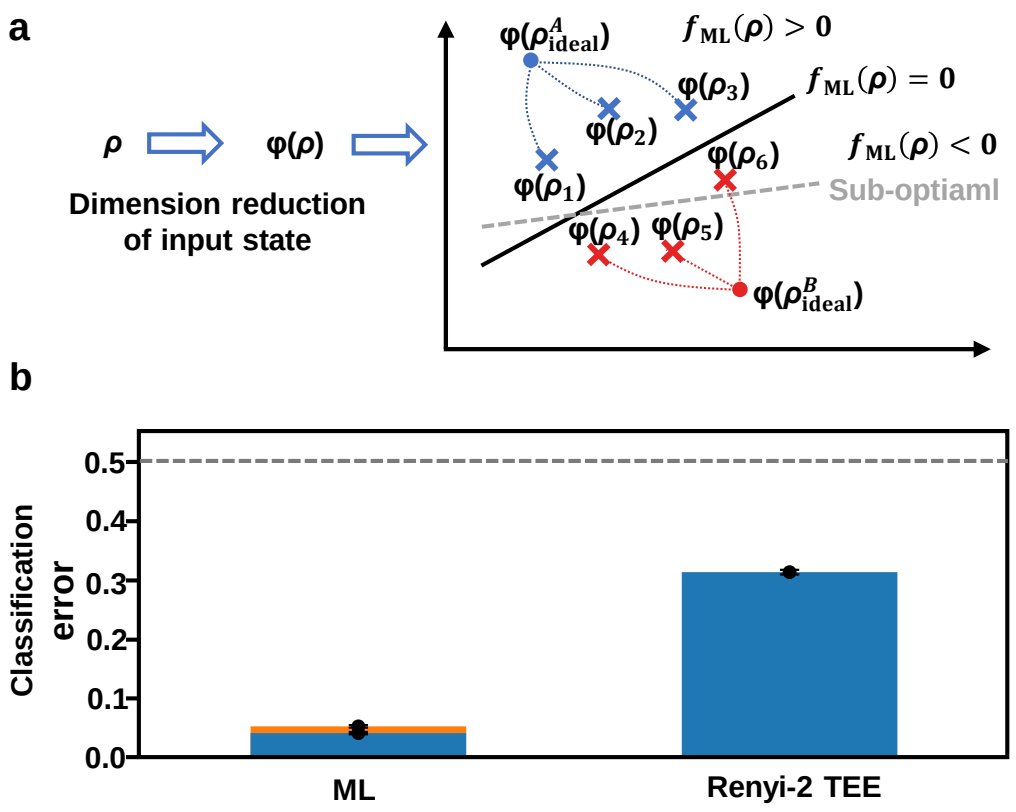

**Fig. 5 | Extracting a phase classifier from a trained ML model. a**, Illustrative diagram of ML procedure. In the first step, given a quantum state $\rho$, we used feature mapping $\varphi(\rho)$ to reduce the dimensionality of the data. To achieve an error-resilient phase classifier $f_{\mathrm{ML}}(\rho)$, training data are generated by applying random unitary transformations to fixed-point states. **b**, Classification error. The orange bars represent the results from the ML model trained on raw data without measurement error mitigation (MEM). The blue bars represent the classification error of the ML model trained by the data with MEM and the Renyi-2 Topological entanglement entropy (TEE). The error bar denotes the standard deviation of 100 instances of the test data set.

# Machine learning on quantum experimental data toward solving quantum many-body problems

Gyungmin Cho and Dohun Kim*

*Department of Physics and Astronomy, and Institute of Applied Physics, Seoul National University, Seoul 08826, Korea*

**Corresponding author: dohunkim@snu.ac.kr*

**Supplementary Information (SI)**

Contents

## Supplementary Note 1: Device characteristics and a qubit selection

We conducted experiments using fixed-frequency superconducting transmon qubits available through IBM cloud and employed a set of gates {$R_Z(\theta)$ ($\theta \in [0, 2\pi]$), SX (= $R_X(\pi/2)$ up to global phase), X (= $R_X(\pi)$ up to global phase)} to implement arbitrary single qubit gates. The native two qubit gate that can be implemented in the IBM hardware is the cross-resonance (CR) gate, and in the experiment, we used echoed cross-resonance (ECR) pulses[1] to reduce gate errors. We utilized the pre-calibrated ECR($\pi$/2) as the basic building block for the two qubits gate. By combining ECR($\pi$/2) and single qubit gates, we can implement CNOT gates (Supplementary Figure S1), allowing us to create universal gates along with arbitrary single qubit gates. The metrics for each qubit and gate were summarized in Supplementary Table 1.

| | |
|---|---|
| $T_1$ | 277.91 us |
| $T_2$ | 166.15 us |
| $\sqrt{X}$(= SX) | $2.262\times10^{-4}$ |
| X | $2.262\times10^{-4}$ |
| ECR(π/2) | $7.476\times10^{-3}$ |
| Readout error | $1.090\times10^{-2}$ |

**Supplementary Table 1** | *ibm_sherbrooke*. Median of the $T_1$, $T_2$, gates (SX, X, ECR(π/2)) error and readout error.

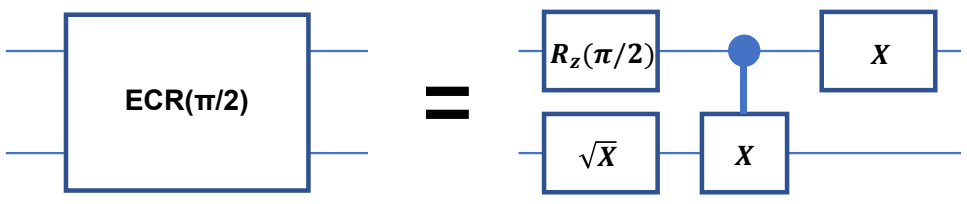

**Supplementary Figure S1** | ECR(π/2) gate decompositions into hardware native gate set.

In the experiment, we chose a qubit layout with the minimum errors by going through a selection process. Due to the constraints in qubit connectivity, we first identified possible layouts capable of running the given circuit. Next, we used device information about the gates and qubits to estimate the total error for each possible layout. Following this, we conducted the actual experiments using the qubit layout that had the lowest error estimate. Since the device characteristics tend to drift over time, we ran experiments on the layout predicted to perform best at the time of the experiment, and the qubits employed in each experiment can be found in Supplementary Note 5. We used the software package mapomatic[2] and Qiskit[3] for this task.

## Supplementary Note 2: Classical shadow and its applications

### 2.1. Classical shadow

To convert the quantum state into a classical form, quantum state tomography (QST) has been used. For QST[4], Pauli basis measurements were usually employed, which have been executed on a variety of quantum hardware platforms[5,6]. Although generalized measurements or even coherent measurements on multiple copies still have an exponential scaling in sample complexity for QST, entire information of the state $\rho$ may not be necessary if we are interested in the expectation value of a local operator $O$. Inspired by this, a state learning method called classical shadow was introduced and implemented experimentally[7,8]. To obtain the classical shadow of a quantum state, we apply a unitary transformation sampled from a specific random unitary ensemble to the state, and then measure the resulting state. In our experiment, we used $\mathrm{Haar}(2)^{\otimes n}$ as the random unitary ensemble. For example, after performing $T$ samplings of random unitary and measurements, we can obtain the experimental results $S_T(\rho)=\{b^{(t)}, U^{(t)}\}_{t=1}^{T}$, $(b^{(t)} \in \{0,1\}^n, U^{(t)}=\otimes_{i=1}^{n} U_i^{(t)})$ where $U_i^{(t)}$ is sampled from the Haar measure over $\mathbb{U}(2)$, from which the empirical average for the quantum state $\rho$ can be written as follows

$$\hat{\sigma}_T(\rho)=1/T\sum_{t=1}^{T}\hat{\rho}_t$$

$$\text{where}\quad \hat{\rho}_t=\otimes_{i=1}^{n}(3U_i^{(t)\dagger}\left|b_i^{(t)}\right\rangle\left\langle b_i^{(t)}\right|U_i^{(t)}-I_2) \tag{SE1}$$

$I_2$ is a 2 by 2 identity matrix. Then, the following facts hold.

**Fact 1** Unbiased estimator[7]. *Define* $\hat{o}=Tr(O\hat{\rho})$ *and use single-shot estimator* $\hat{\rho}=\otimes_{i=1}^{n}(3U_i^{\dagger}|b_i\rangle\langle b_i|U_i-I_2)$ *where each* $U_i$ *is sampled from Unitary 2-design over* $\mathbb{U}(2)$*, then expectation value of* $\hat{o}$ *gives*

$$\mathbb{E}_{U,b}(\hat{o})=\mathrm{Tr}(O\rho) \tag{SE2}$$

And the associated variance is given by the following.

**Fact 2** Upper bound on the variance of estimation[7]. *Define* $\hat{o} = Tr(O\hat{\rho})$ *and use single-shot estimator* $\hat{\rho} = \otimes_{i=1}^{n}(3U_i^\dagger |b_i\rangle\langle b_i| U_i - I_2)$ *where each* $U_i$ *is sampled from Unitary 3-design over* $\mathbb{U}(2)$*, then upper bound of the variance of* $\hat{o}$ *obeys*

$$\mathrm{Var}(\hat{o}) \le 2^{locality(O)} \|O\|_2^2 \le 4^{locality(O)} \|O\|_\infty^2 \tag{SE3}$$

where locality($O$) is the number of qubits where operator $O$ acts non-trivially. So, using empirical average $\hat{\sigma}_T(\rho)$ of (1) in the main text, we can reduce the variance by increasing the number of experimental trials ($T$). In experiments, we used Haar(2)$^{\otimes n}$ as a random unitary ensemble, so the above two facts automatically satisfied.

Using the Fact1 and 2, we can prove the following Corollary1.

**Corollary1** Virtual gate by post-processing on the classical shadows. *If unitary V can be decomposed into tensor products of single qubits gate (* $V = \otimes_{i=1}^{n} V_i$ *), Using the classical shadows* $S_T(\rho) = \{b^{(t)}, U^{(t)}\}_{t=1}^{T}$*, we can obtain unbiased estimator* $\hat{o}_V = \mathrm{Tr}(OV\hat{\rho}V^\dagger)$ *with the same upper bound on the variance as in Fact 2.*

*Proof.* we can directly use the result from Fact 1, 2.

$$\begin{aligned}\mathbb{E}(\hat{o}_V) &= \mathbb{E}(\mathrm{Tr}(OV\hat{\rho}V^\dagger)) \\ &= \mathrm{Tr}(OV\mathbb{E}(\hat{\rho})V^\dagger) \quad (\because \text{linearity of } \mathrm{Tr}(\cdot)) \\ &= \mathrm{Tr}(OV\rho V^\dagger) \quad (\because \text{ Fact 1})\end{aligned} \tag{SE4}$$

As expected, it gives an unbiased result. And we can obtain upper bound on the variance.

$$\begin{aligned}\mathrm{Var}(\hat{o}_V) &\le 2^{\mathrm{locality}(V^\dagger O V)} \|V^\dagger O V\|_2^2 \quad (\because \mathrm{Tr}(OV\hat{\rho}V^\dagger) = \mathrm{Tr}(V^\dagger O V\hat{\rho})) \\ &= 2^{\mathrm{locality}(O)} \|O\|_2^2\end{aligned} \tag{SE5}$$

In the second line, we use the fact that tensor products of single qubit gates do not change the locality of $O$ and matrix norm $\|\cdot\|_2$ is invariant under the unitary transformation. □

## Numerical simulation for virtual gate

We provide a numerical simulation of the virtual gate method for estimating the expectation value for a given quantum state. In our simulation, we prepared a 5-qubit GHZ state with a random unitary applied and tried to estimate the expectation values of $\{Z_i Z_{i+1}\}_i$ via the classical shadow. As shown in Supplementary Figure S2a, the estimated values obtained from post-processing on classical shadows of GHZ state and the classical shadow from directly applying the random unitary both yield the same unbiased results. Additionally, both methods exhibit the same scaling of standard deviation (std) as the number of experimental trials ($T$) increases, as depicted in Supplementary Figure S2b. For the circuit simulation, we used the software package Qiskit[3].

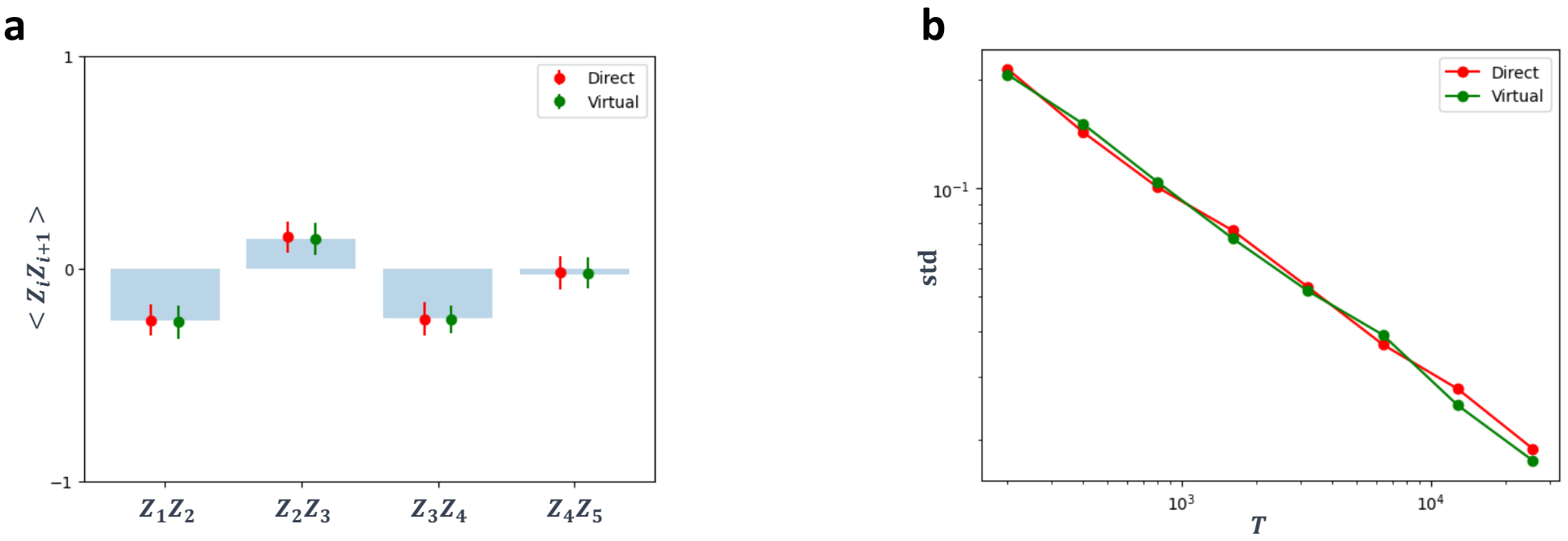

**Supplementary Figure S2** | Classical simulation of virtual gate method. **a**, blue bars represent the exact values for each observable $Z_i Z_{i+1}$, the red ones denote the estimated values using the classical shadow obtained through $T = 1600$ randomized measurements (RM) after actually applying the tensor products of single qubit gates, and the green ones are the values estimated through virtual gate method. The error bar is the standard deviation (std) from 100 data points. **b**, the scaling behavior of standard deviation on the number of randomized measurements ($T$).

## 2.2. Shadow kernel

In our quantum phase classification experiment, we employed the shadow kernel (3) in the main text to learn the non-linear properties of the state $\rho$. If we choose the kernel function $k_{\text{linear}}(x, x') = \text{Tr}(\rho(x)\rho(x'))$ for the task of distinguishing quantum phase where $\rho(x)$ is the ground state of $H(x)$, then there must exist a function $f(\rho) = \text{Tr}(O\rho)$ that can be used for phase classification, such that $f(\rho) \geq 0$ if $\rho \in$ phase $A$ and $f(\rho) < 0$ if $\rho \in$ phase $B$. However, recent studies have indicated that using such linear observables is likely to fail when the system is not translational invariant[9,10]. As a result, a more expressive kernel is needed to perform phase classification in more general setting, prompting us to use the shadow kernel for ML. If a function $f(\rho)$ for some non-negative integer $r$ and $d_p$

$$\boldsymbol{f}(\rho) = \sum_{\boldsymbol{d}=0}^{d_p} 1/\boldsymbol{d}!\sum_{A_1 \dots A_d \subset \{1,\dots,\boldsymbol{n}\},|A_i| \leq \boldsymbol{r}} \text{Tr}(\boldsymbol{O}_{A_1 \dots A_d} \text{Tr}_{\neg A_1}(\rho) \otimes \cdots \otimes \text{Tr}_{\neg A_d}(\rho)) \tag{SE6}$$

exists that can classify the given phases, then the shadow kernel can be used to learn the function $f(\rho)$. Furthermore, if the system's correlation length is independent on the system size, it has been proven that learning $f(\rho)$ can be done using only polynomial sample and computation time[10]. During the ML in our case, we utilized a modified shadow kernel (SE7) for numerical stability, rather than the original shadow kernel (3) in the main text.

$$k_{\text{shadow}}(S_T(\rho),\ S_T(\tilde{\rho})) = \exp\{\tau/T(T-1)\sum_{t \neq t'}^{T} \exp[\gamma/n\sum_{i=1}^{n} \text{Tr}(\sigma_i^{(t)}\tilde{\sigma}_i^{(t')})]\}$$

$$\text{where } \sigma_i^{(t)} = 3U_i^{(t)\dagger} \left|b_i^{(t)}\right\rangle\left\langle b_i^{(t)}\right| U_i^{(t)} - I_2 \tag{SE7}$$

**Supplementary Note 3: Various error reducing procedures**

We conducted classical ML using data from a quantum computer without quantum error correction, which means that various types of errors exist in the data. If there are a high level of errors in the data, it becomes difficult to obtain accurate predictions from the trained ML model even with using effective ML algorithms. Therefore, to achieve better performance of ML model, it is imperative to obtain data with reduced errors. In this Note, we will introduce various error reducing procedures used in experiments.

3.1. Dynamical Decoupling (DD)

During the experiment in the main text, the gate time for single-qubit and two-qubit gates are different, and there are cases where gates are intentionally not applied to some qubits when running circuits. As a result, while operations are performed on some specific qubits, the remaining qubits are idle, leading to dephasing errors. DD is a technique to counteract these errors by inserting a logical identity sequence during the idle period, causing coherent errors to refocus[11,12]. Among many possible pulse sequences, we utilized [X(+$\pi$), Y(+$\pi$), X(-$\pi$), Y(-$\pi$)] during the idle period in the all experiments of main text (In cases where the duration is insufficient to implement the above sequence, [X(+$\pi$), X(-$\pi$)] is utilized instead).

3.2.Pauli Twirling (PT)

While DD is applied to idle qubits, PT consists of adding a pair of Pauli gates before and after the two-qubit Clifford gates which doesn't change the logical outcome[13]. This allows us to transform some coherent errors in two-qubit gate into random decoherent errors. Given that, in the worst case, coherent errors can accumulate quadratically with circuit depth compared to decoherent errors, PT can help decrease the errors in the experiment. The native two-qubit gate employed in the experiment is ECR($\pi$/2) which is in the Clifford group. We implemented PT by uniformly sampling a Pauli gate $P \in$ {II, IX…ZY, ZZ} in front of each ECR($\pi$/2) gate and inserting another Pauli gate $P' = \mathrm{ECR}(\pi/2)\cdot P\cdot \mathrm{ECR}(\pi/2)^\dagger$ afterward to compensate the logical effect of the $P$. Because altering the gate sequence for each measurement would be resource-intensive, we used 5 Pauli-twirled circuit instances in each circuit (But, when acquiring classical shadows, PT was implemented for all $T$ randomized measurements).

3.3. Post-selection (PS)

To obtain data with fewer errors, one approach is to reduce the errors themselves. Alternatively, post-processing methods that detect errors and discard the erroneous data can also be used to achieve fewer-error data, and PS falls under this category. For instance, in a second-quantized system with a conserved number of particles, where $|0\rangle$ means the absence of particles and $|1\rangle$ the presence of the particle. After the measurement, If the total number of particles is not conserved, it indicates errors occurred during the state preparation or measurements. By excluding this data during the analysis, results with reduced errors can be acquired at the cost of more measurements than before.

In the experiment of main text, PS was used in estimating site correlation $\langle a_i^\dagger a_j\rangle$ for the ground state of a 1D nearest-neighbor random hopping system. The ground state of the given system should be half-filled, so in a system with $n$ sites, only $n/2$ ($n$ = even) sites was occupied. If measurement outcomes give different particle number, we can omit that result. Such an approach can be utilized in broader scenarios where the parity is conserved[12].

3.4. Macweeny-purification

In the experiment of predicting ground state properties, our goal was to learn the correlation matrix $C$ where $C_{ij}$ corresponds to site correlation $\langle a_i^\dagger a_j \rangle$ of the ground state. In cases like the 1D nearest-neighbor random hopping system, where the system's ground state is expressed as a Slater determinant, the correlation matrix becomes idempotent ($C^2 = C$). However, in actual experiments, errors occurred during quantum state preparation or measurements and statistical fluctuation of measurement outcomes cause correlation matrix $C_{\mathrm{exp}}$ to deviate from idempotency. Therefore, a process of projecting it onto an idempotent matrix is helpful, which can be implemented using Macweeny-purification[14]. Macweeny-purification involves iteratively repeating the formula (SE8).

$$C_{k+1} = C_k^2(3I - 2C_k) \tag{SE8}$$

The correlation matrices drawn in Fig. 2 of the main text correspond to the purified version of the raw correlation matrix $C_{\mathrm{exp}}$ and $C_{\mathrm{ML}}$. A detailed error analysis of Macweeny-purification can be found in the literature[14].

3.5. Parity measurement via circuit recompiling

We were able to eliminate some experimental results through post-selection (PS), but in order to do this, the circuit must respect a certain symmetry. For example, in the experiment for predicting ground state properties described in the main text, the applied circuit should not change the number of particles to allow post-selection to proceed.

The quantity $a_i^\dagger a_j$ that we want to measure in the experiment is not Hermitian when $i \neq j$. But, $\langle a_i^\dagger a_j \rangle$ can be obtained by combining the measurement results of $\langle a_i^\dagger a_j + a_j^\dagger a_i \rangle$ and $\langle \mathrm{i}(a_i^\dagger a_j - a_j^\dagger a_i) \rangle$ which are Hermitian. However, in our case, the unitary and initial states of the circuit are all composed of real values, so $\langle a_i^\dagger a_j\rangle = 2\cdot\mathrm{Re}(\langle a_i^\dagger a_j + a_j^\dagger a_i\rangle)$, and it is transformed into $1/4(X_iZ_{i+1}..Z_{j-1}X_j + Y_iZ_{i+1}..Z_{j-1}Y_j)$ by the Jordan Wigner transformation. First, in the case where $i = j$, $a_i^\dagger a_i$ is transformed into $1/2(1 - Z_i)$ in which computational basis measurement is enough. The second case where $j = i+1$, we need to measure $1/4(X_iX_{i+1} + Y_iY_{i+1})$, and since $X_iX_{i+1}$ and $Y_iY_{i+1}$ are commute, we can simultaneously measure them by rotating the measurement basis employing $U_{\mathrm{single}}^{\otimes 2}$ where $U_{\mathrm{single}} = \mathrm{R}_X(\pi/2)\cdot\mathrm{R}_Y(-\pi/2)$ before the measurement. However, the $U_{\mathrm{single}}$ does not preserve the particle number, making post-selection impossible for subsequent measurement results. To circumvent this issue, previous research[12,14] has shown that applying $\mathcal{U}_{\mathrm{p}}^{(\mathrm{layer})}$ (consisting of a layer of two qubit gates preserving the symmetry) instead of $U_{\mathrm{single}}$ allows for post-selection in measurement results. We further improved this by absorbing the two-qubit gates in $\mathcal{U}_{\mathrm{p}}^{(\mathrm{layer})}$ into the gates required for the state preparation circuit, without incurring any extra gate overhead. In the following, we will provide a detailed derivation of parity measurement via circuit recompiling. To do that, we use some facts.

**Fact 3** Basis transformation of Fermionic creation/annihilation operator[15]. *Consider* $A_K = \sum_{i,j} K_{ij} c_i^\dagger c_j$ *where* $K$ *is* $N \times N$ *Hermitian matrix and* $\mathbf{c}^\dagger = (c_1^\dagger \ldots c_N^\dagger)^T$*. Then, the following holds*

$$\exp(-iA_K)\mathbf{c}^\dagger \exp(iA_K) = K\mathbf{c}^\dagger \tag{SE9}$$

Using the Fact 3 we can define $\mathcal{U}(U) = \exp(\sum_{ij}(\log U)_{ij}^T c_i^\dagger c_j)$ where $T$ is transpose, then (SE9) change into the following

$$\mathcal{U}(U)\mathbf{c}^\dagger \mathcal{U}(U)^\dagger = U\mathbf{c}^\dagger \tag{SE10}$$

Using the (SE10), it becomes clear that $\mathcal{U}(U_1U_2) = \mathcal{U}(U_2)\mathcal{U}(U_1)$ holds.

**Fact 4** Givens decompositions[15]. *$M \times N$ ($M \leq N$) matrix $Q$ satisfying $Q^{\dagger}Q = P_S$ where $P_S$ is the projector onto suitable subspace S which has dimension M can be decomposed iteratively into $N \times N$ Givens matrices (G) and a $M \times N$ diagonal matrix.*

$$Q = U_{\text{diag}} G_{N_G} \ldots G_1 \text{ where } G_i(\theta) = \begin{pmatrix} 1 & & & & & & & \\ & \ldots & & & & & & \\ & & 1 & & & & & \\ & & & \cos\theta & -\sin\theta & & & \\ & & & \sin\theta & \cos\theta & & & \\ & & & & & 1 & & \\ & & & & & & \ldots & \\ & & & & & & & 1 \end{pmatrix} \tag{SE11}$$

$N_G$ is the total number of Givens matrices. The Givens matrices ($G$) used in this paper are non-trivial only within two neighboring rows and columns. Keep that in mind, when we compute $\mathcal{G} = \mathcal{U}(G)$, it effectively transforms into a two-qubit gate acting on adjacent qubits (SE12).

$$\mathcal{G} = \mathcal{U}(G(\theta)) = \begin{pmatrix} 1 & 0 & 0 & 0 \\ 0 & \cos\theta & -\sin\theta & 0 \\ 0 & \sin\theta & \cos\theta & 0 \\ 0 & 0 & 0 & 1 \end{pmatrix} \tag{SE12}$$

**Fact5** Converting a quadratic Hamiltonian into a non-interacting Hamiltonian[15]. *The quadratic Hamiltonian written as* $H = \sum_{i,j=1}^{N} (K_{ij} - \mu\delta_{ij}) c_i^{\dagger} c_j$ *where $K = K^{\dagger}$ and $\mu$ is the chemical potential can be converted into* $H = \sum_i \varepsilon_i b_i^{\dagger} b_i + (c \text{ number})$ *where $\boldsymbol{b}^{\dagger} = U\boldsymbol{c}^{\dagger}$. As a result, ground state of H can be written as* $|gs\rangle = \prod_{i,\,\varepsilon_i<0} b_i^{\dagger} |\text{vac}_b\rangle = \mathcal{U}(U) \prod_{i,\,\varepsilon_i<0} c_i^{\dagger} |\text{vac}_c\rangle,\ (^{\forall} i,\ b(c)_i |\text{vac}_{b(c)}\rangle = 0)$.

In experiments for predicting ground state properties, we use the Hamiltonian $H_{\text{hop}} = \sum_i x_i c_i^{\dagger} c_{i+1} + h.c.$ which is a special case of quadratic Hamiltonian. Utilizing Fact 5, we can transform $H_{\text{hop}}$ into a non-interacting Hamiltonian $H_{\text{hop}} = \sum_i \varepsilon_i b_i^{\dagger} b_i + (c \text{ number})$ where $\boldsymbol{b}^{\dagger}$ =

$U\mathbf{c}^\dagger$. By keeping $i^{\text{th}}$-row where $\varepsilon_i < 0$, we can obtain the $M \times N$ matrix $Q$ ($Q^\dagger Q = P_S$) from $U$. Using Fact 3 and 4, by applying the unitary $\mathcal{U}$

$$\begin{aligned}\mathcal{U} &= \mathcal{U}(G_{N_G}...G_1) \\ &= \mathcal{U}(G_1)...\mathcal{U}(G_{N_G})\end{aligned} \tag{SE13}$$

to the half-filled $|1...10...0\rangle$ state, we can prepare the ground state up to a global phase.

In the experiment, in addition to $\mathcal{U}$ in (SE13), a basis rotation $\mathcal{U}_{\text{p}}^{(\text{layer})}$ is needed for parity measurement. Taking this into account, the total unitary to be implemented through a quantum computer is $\mathcal{U}_{\text{total}} = \mathcal{U}_{\text{p}}^{(\text{layer})} \cdot \mathcal{U}$. Each two-qubit gate in $\mathcal{U}_{\text{p}}^{(\text{layer})}$ can be obtained by diagonalizing $1/4(X_iX_{i+1} + Y_iY_{i+1})$.

$$1/4(X_iX_{i+1} + Y_iY_{i+1}) = U_{\text{p}} D U_{\text{p}}^\dagger \quad \text{where} \quad U_{\text{p}} = \begin{pmatrix} 1 & 0 & 0 & 0 \\ 0 & 1/\sqrt{2} & 1/\sqrt{2} & 0 \\ 0 & -1/\sqrt{2} & 1/\sqrt{2} & 0 \\ 0 & 0 & 0 & 1 \end{pmatrix},\ D = \begin{pmatrix} 1 & 0 & 0 & 0 \\ 0 & 1/2 & 0 & 0 \\ 0 & 0 & -1/2 & 0 \\ 0 & 0 & 0 & 1 \end{pmatrix} \tag{SE14}$$

By utilizing (SE14), $1/4\,\text{Tr}((X_iX_{i+1} + Y_iY_{i+1})\rho) = \text{Tr}(U_{\text{p}} D U_{\text{p}}^\dagger \rho) = \text{Tr}(D U_{\text{p}}^\dagger \rho U_{\text{p}})$ holds. As a result, $U_{\text{p}}^\dagger$ is adequate for the parity measurement. However, considering that the error of two-qubit gates is an order of magnitude larger than single-qubit gate, additional errors from $U_{\text{p}}^\dagger$ in $\mathcal{U}_{\text{p}}^{(\text{layer})}$ occurs. Yet, taking advantage of $U_{\text{p}}^\dagger = \mathcal{U}(\text{G}(\pi/4))$, we are able to accomplish parity measurement without any additional two-qubit gates through the following circuit recompiling.

$$\begin{aligned}\mathcal{U}_{\text{total}} &= \mathcal{U}_{\text{p}}^{(\text{layer})} \mathcal{U}(G_{N_G}...G_1) \\ &= \mathcal{U}(P)\mathcal{U}(G_1)...\mathcal{U}(G_{N_G}) \\ &= \mathcal{U}(G_{N_G}...G_1 P) \\ &= \mathcal{U}(G'_{N_G}...G'_1) \\ &= \mathcal{U}(G'_1)...\mathcal{U}(G'_{N_G})\end{aligned} \tag{SE15}$$

Through this, we achieved the desired $\mathcal{U}_{\text{total}}$ with a new set of Givens rotations $\{G_i'\}$. By adopting the circuit recompiling method outlined above, we can eliminate the errors from parity measurement. In Supplementary Figure S3, we conducted a noisy classical simulation while varying the degree of depolarizing error rate and observed that the total error decreased in all depolarizing strength.

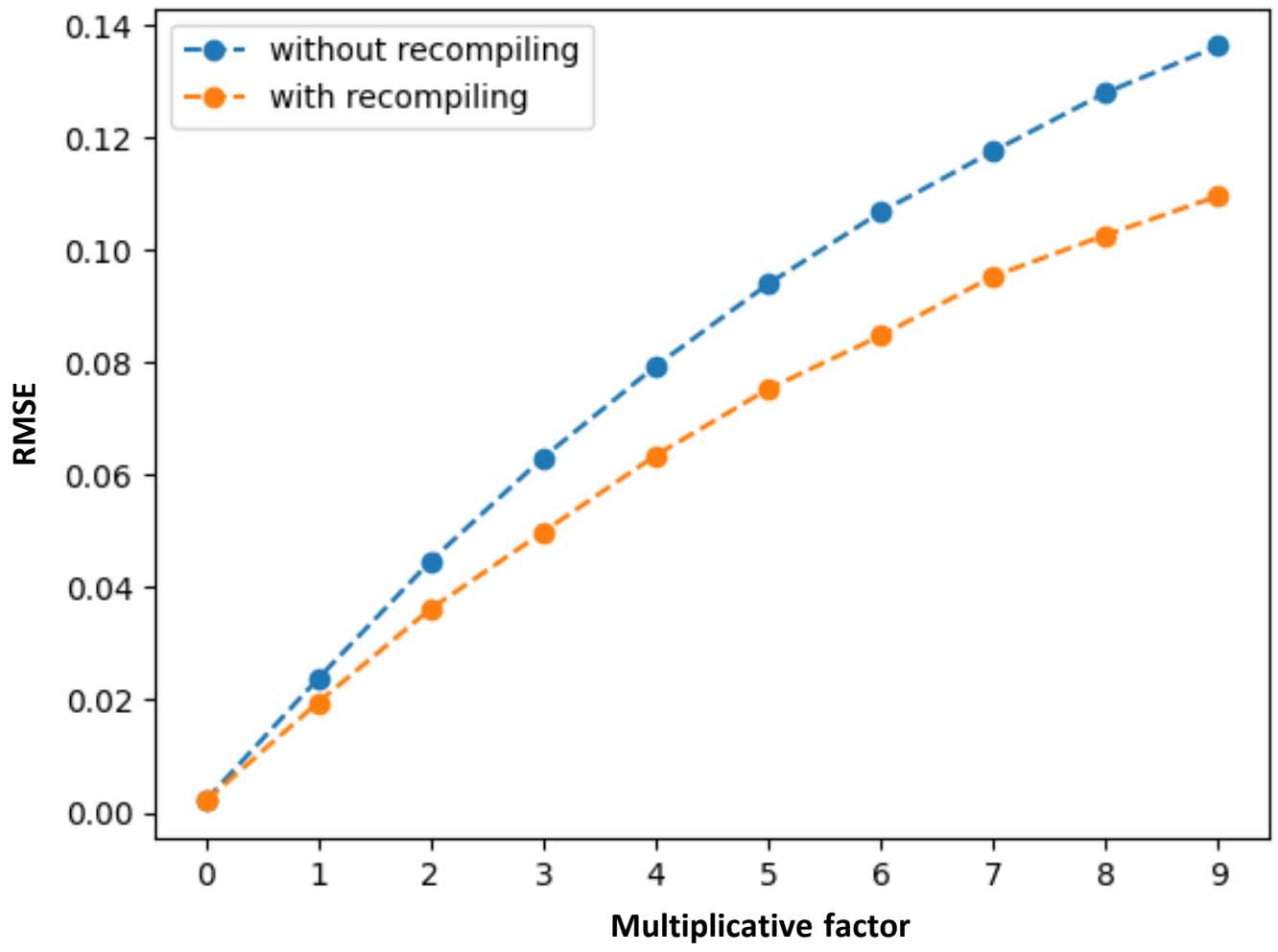

**Supplementary Figure S3** | Simulation of 6 qubits random hopping system to determine the degree to which parity measurement via circuit recompiling can reduce errors. The blue points indicate the actual implementation of $\mathcal{U}_{\text{p}}^{(\text{layer})}$, and the orange points depict the outcomes with circuit recompiling. In the noisy simulation, we imposed the depolarizing channel with error rate ($p_{\text{single}}$, $p_{\text{two}}$) = (0.001, 0.01) × (multiplicative factor) on each single-qubit and two-qubit gate.

Until now, we have focused on a specific case where $j = i+1$. However, for a more general case where $i \neq j$, it's possible to transform $a_i^\dagger a_j$, using fermionic swap operations, to the form of $a_k^\dagger a_{k+1}$ without additional gates by adjusting angles in Givens rotations[14].

### 3.6. Measurement-assisted state preparation method

Recently, various methods have been proposed and experimentally realized to prepare topologically ordered states on the quantum computer[16]. However, to implement this, suitable qubit connectivity needs to meet, and if this condition is not satisfied, swap gate overhead will occur. Moreover, even if the qubit connectivity condition was met, to prepare the ground state of the toric code using only unitary gates, a required circuit depth is proportional to the code distance ($d_{\mathrm{code}}$) of toric code.

To overcome such problems, we used the Measurement-assisted state preparation method as a way to detour hardware constraints and reduce errors from additional swap gates[17]. As shown in Fig. 4a in the main text, after applying a series of unitary operations and measuring ancilla qubits, the state after the measurement can be written as follows.

$$|\psi\rangle = 1/\sqrt{2^{N_p}} \prod_p (I + (-1)^{m_p} B_p)|0\rangle^{\otimes n} \tag{SE16}$$

where $B_p = \prod_{i\in p} X_i$ is a $X$-plaquette operator, $N_p$ is the number of $X$-plaquettes and $m_p \in \{0, 1\}$ represents the measurement results of the ancilla qubit present on each $X$-plaquette.

To create the $|0_L\rangle$ state corresponding to (4) in the main text, we need to apply adaptive gates based on the measurement results to ensure all $m_p$ values are 0 in (SE16). In our experiment, by applying the Pauli Z gate depending on the measurement results of the ancilla qubits, we were able to deterministically prepare the $|0_L\rangle$ state. We grouped some of the 25 data qubits as shown in the Supplementary Figure S4 and applied the adaptive Pauli Z gates only within the grouped qubits.

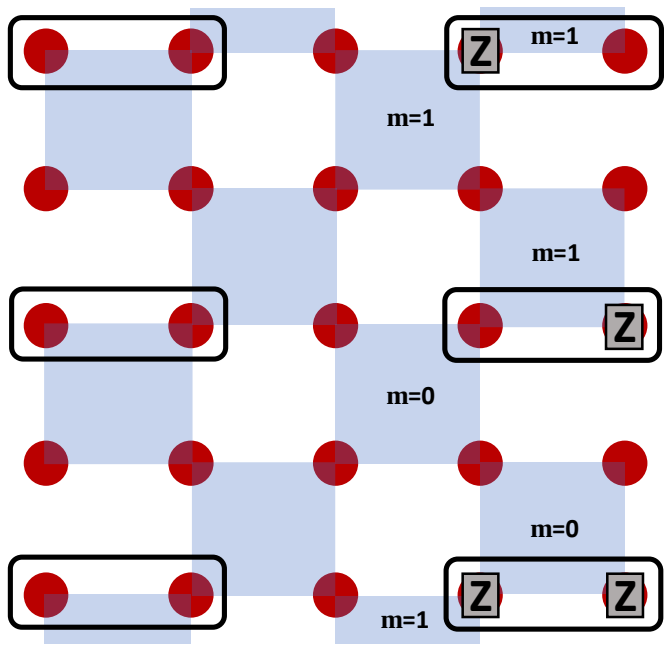

**Supplementary Figure S4 |** The colored with blue squares corresponds to the $X$-plaquette, and there is a measurement result for each ancilla qubit on the $X$-plaquette. Among the many possible ways, we used the one

that applies suitable Z gates to qubits grouped together by black lines.

In Fig. 4a of the main text, 8 CNOT gates were used to realize the projection $\frac{I+(-1)^{m_p}B_p}{2}$. Also, in the heavy-hexagonal lattice, in addition to 1 measurement qubit and 4 data qubits, 2 additional ancilla qubits initialized to $|0\rangle$ are required to apply the CNOT gate from the measurement qubit to the data qubit as in the Supplementary Figure S5.

**Supplementary Figure S5** | 4 additional CNOT gate due to the qubit connectivity on heavy-hexagonal lattice

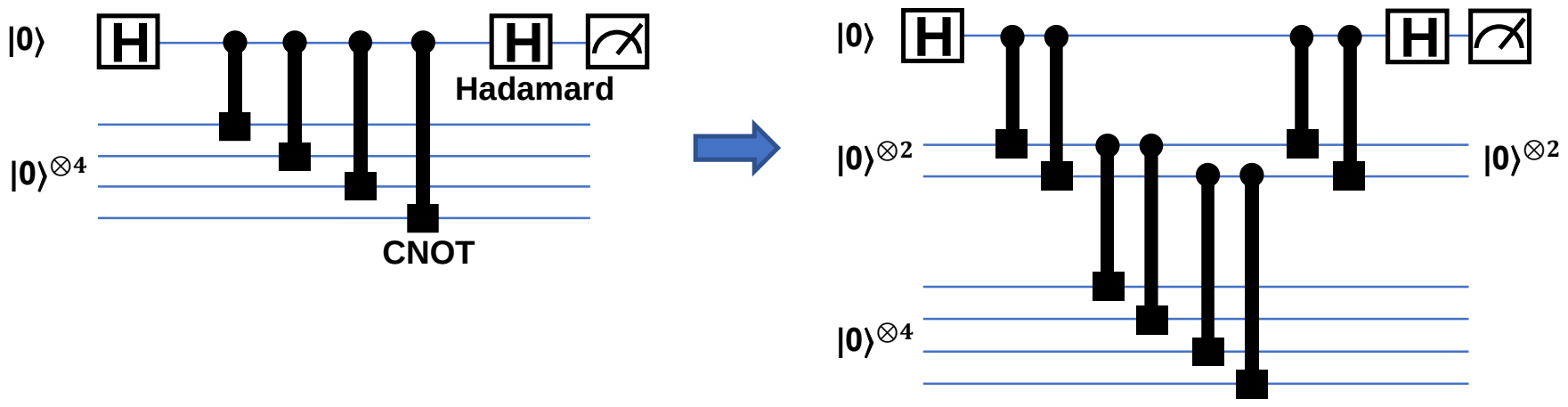

3.7. Virtual gate on classical shadows

During the measurement of ancilla qubits in the Measurement-assisted state preparation method, data qubits remain idle, and considering that the measurement time is much longer than the single and two-qubit gate time in most superconducting devices, unwanted errors may occur during this period, causing the information stored in the qubits to leak out. To reduce the errors from the idle data qubits, we used a virtual gate through post-processing on classical shadows, as explained in the Supplementary Note 2.

To create training data for the topologically ordered phase, we apply a tensor product of random unitary $U = \bigotimes_{i=1} U_i$ (each $U_i$ is sampled independently from Haar(2)) to the fixed-point state $|0_L\rangle$, and perform randomized measurements to obtain the corresponding classical shadows for each state. It is important to note that there are three types of randomness involved here. The first one is for data generation, the second one is the random unitary required to obtain the classical shadow representation of the state, and the last is adaptive gate conditioned on the measurement results of ancilla qubits. In experiments on the quantum hardware, we only applied the random unitary for acquiring classical shadows representation, while the random unitary required for data generation and adaptive gate are done via post-processing on classical shadows.

## 3.8.Measurement error mitigation (MEM)

During the experiment for Fig. 5 in the main text, we used MEM to correct the measurement errors in the data obtained from the quantum computer. We carried out an additional experiment to construct the response matrix ($R$) containing information about the correlation of errors that occurred during the measurements. Since we conducted measurements on 4 out of the 9 qubits in the experiment, we prepared possible 16 strings for the 4 qubits and obtained response matrix $R$ through 50,000 measurements for each string. Using the acquired $R$, we could retrieve the error-mitigated measurement outcome $p_{\mathrm{MEM}}$ by employing the inverse relation $p_{\mathrm{MEM}} = R^{-1}p_{\mathrm{exp}}$.

## Supplementary Note 4: Comparison with quantum convolutional neural network (QCNN)

Following the development of QCNN[18], which was inspired by the well-known convolutional neural network (CNN) architecture in classical ML[19], various subsequent studies have been conducted[9,20,21]. There are two main reasons for the success of QCNN. First, while it was inspired by classical ML, the physical roots of QCNN can be found in the well-known multi-scaled entanglement renormalization ansatz (MERA)[22] and tensor renormalization[23] from tensor networks, allowing the ML model to be more than just a simple black box and providing physical interpretability. Secondly, even though many circuit ansätze in quantum machine learning suffer from the barren plateau[24], it is known that QCNN ansatz doesn't[20]. Building on this, experimental realization has shown that even a restricted QCNN compared with original proposal can discern SPT phases with enhanced accuracy than naïve measurements of the string order parameter (SOP) on a 7-qubit superconducting device[21].

Recently, further research has explored the performance of QCNNs in situations without translational invariance[9]. Despite the given state being in the SPT phase, they reported that the performance of the QCNN decreases as the system's non-uniformity increases. QCNNs for distinguishing the SPT phase are based on the SOP and in the process of deriving the SOP, translational invariance of system was assumed[25]. Therefore, in cases without translational invariance, the performance of phase classification through the SOP cannot be guaranteed. Furthermore, in some cases, for linear order parameters $f_{\text{linear}}(\rho) = \mathrm{Tr}(O\rho)$, detecting a specific SPT phase is proven impossible, even $O$ spans the entire system[9]. To circumvent this no-go theorem, using $f_{\text{non-linear}}(\rho) = (O\rho^{\otimes k})$ for phase classification is one approach, exemplified by many-body topological invariants (MBTIs)[26].

Experimentally measuring non-linear properties on noisy quantum computers is challenging, but recent studies have implemented the use of classical shadow or statistical correlation to measure non-linear quantities such as $\mathrm{Tr}(\rho^2)$ and $\mathrm{Tr}(\rho^3)$[27]. In our case, by using classical shadows as data in ML, we demonstrated that phase classification between the SPT phase and the trivial phase is possible through ML, even in cases where the system lacks translational invariance, making the task challenging for SOPs or QCNN to succeed.

In Fig. 3g of the main text, we demonstrated that the ML model trained on states generated from fixed-points state can also be applied to distinguish the phases of translationally invariant systems, such as $H_{\mathrm{CI}} = -J\sum_i Z_i X_{i+1} Z_{i+2} - h_1 \sum_i X_i - h_2 \sum_i X_i X_{i+1}$ used in previous studies[18,21]. To obtain the classical shadow representation of the ground state of the $H_{\mathrm{CI}}$, we employed the DMRG method. After fixing $J = 1$, we performed sampling of $h_1$ and $h_2$ from [0.0, 1.6] and [-0.8, 1.6]. By mapping these data points to the space with the phase boundary from the trained ML model, we can classify between SPT and trivial phases with high accuracy. Distributions of data mapped on the suitable space are shown in the Supplementary Figure S6.

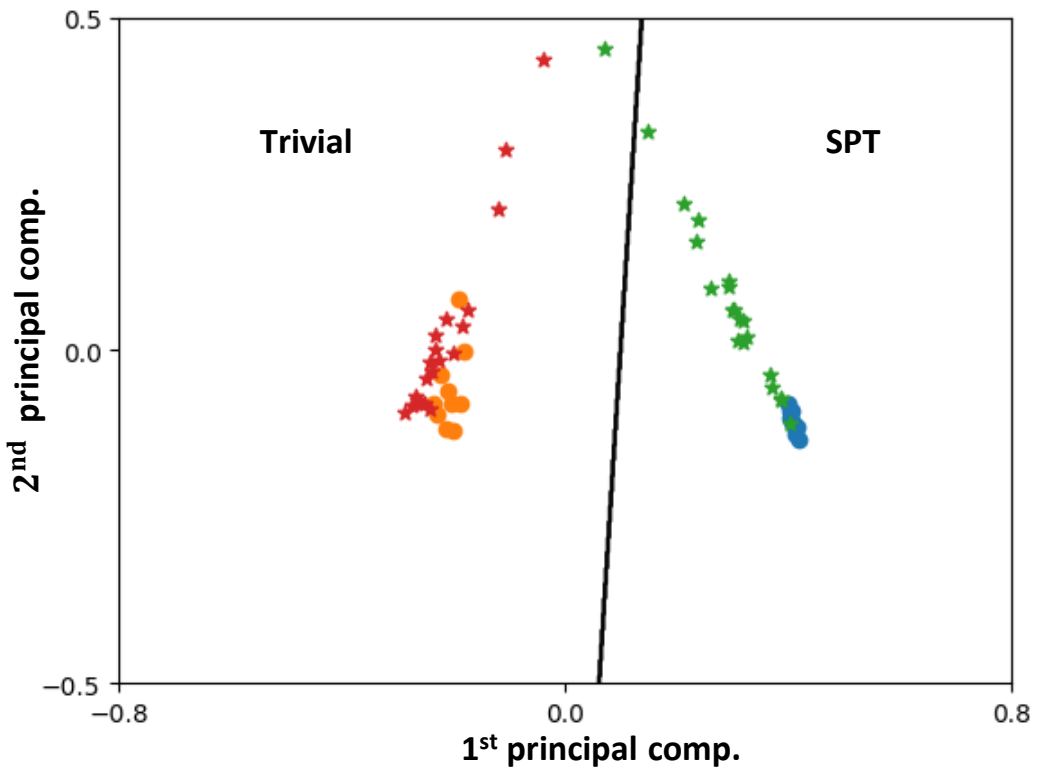

**Supplementary Figure S6 |** Distribution of training and test data. Circle represents training data, and asterisk represents test data. Blue and green represents SPT phase, while orange and red represents trivial phase.

By repeating the above process 10 times, we obtained the success probability of phase classification to be 0.945 on average.

**Supplementary Note 5: Experimental details**

In this Note, we will discuss the details of the experiments and learning processes carried out in the main text.

*Predicting properties of the ground state*

5.1. Learning the ground state properties of 1D nearest-neighbor random hopping system

In our study, we tried to learn the ground state properties of a 1D nearest-neighbor (NN) random hopping system $H_{\text{hop}} = \sum_i x_i (a_i^\dagger a_{i+1} + a_{i+1}^\dagger a_i)$ where $x_i$ was sampled from [0, 2] and, in this parameter regimes, there are multiple quantum phases. In the theoretical proof regarding the performance of ML model[10], they assumed that both training and test data are within the same phase. However, we expanded the learning task to a more realistic and practical scenario (where it's unknown which phases the given system have) and observed that widely used ML models work effectively even this case with favorable scaling $N_{\text{data}} \sim O(1/\text{RMSE})$ where RMSE is Root-Mean-Square error of the ML model.

5.1.1 Kernel regression

In the Fig. 2 of main text, we calculated the value of $f(x_{\text{new}})$ for a new point $x_{\text{new}}$ using the following formula

$$f(x_{\text{new}}) = \sum_{i=1}^{N_{\text{data}}} \sum_{j=1}^{N_{\text{data}}} k(x_{\text{new}}, x_i)(K + \lambda I)_{ij}^{-1} f(x_j) \qquad \text{(SE17)}$$

where $\lambda$ is hyperparameter, $K_{ij} = k(x_i, x_j')$ is a kernel matrix and $I$ is $N_{\text{data}} \times N_{\text{data}}$ identity matrix. Using the kernel trick $k(x, x') = \varphi(x)^T \varphi(x)$, and $w^*$ which minimizes the $L_2$ loss

$$w^* = \underset{w}{\operatorname{argmin}} \sum_{i=1}^{N_{data}} \frac{1}{2} \left| w^T \phi(x_i) - f(x_i) \right|^2 + \frac{\lambda}{2} \|w\|_2^2 \quad \text{(SE18)}$$

where the second term here is used for regularization, the predicted value for $x_{\mathrm{new}}$ can be written as $f_{\mathrm{ML}}(x_{\mathrm{new}}) = w^{*T} \varphi(x_{\mathrm{new}})$, and by simplifying this, (SE17) can be derived[19].

In the experiments, ML was conducted using not only the (modified) Dirichlet kernel but also Gaussian kernel and neural tangent kernel (NTK)[28]. The calculation of the NTK function was performed using the software package neural-tangents[29]. The specific expressions for each kernel function are as follows

$$\tilde{k}(x, x') = \sum_{i \neq j} \sum_{k_i=-3}^{3} \sum_{k_j=-3}^{3} \cos(\pi(k_i(x_i - x_i') + k_j(x_j - x_j'))) \quad \text{(modified) Dirichlet kernel} \quad \text{(SE19a)}$$

$$\tilde{k}(x, x') = \exp(-\alpha \|x - x'\|_2^2) \text{ where } \alpha = N_{data}^2 \Big/ \sum_{i=1}^{N_{data}} \sum_{j=1}^{N_{data}} \|x_i - x_j\|_2^2 \quad \text{Gaussian kernel} \quad \text{(SE19b)}$$

$$\tilde{k}(x, x') = k^{(\mathrm{NTK})}(x, x') \quad \text{NTK} \quad \text{(SE19c)}$$

We normalized $\tilde{k}(x, x')$ into $k(x,x') = \tilde{k}(x,x') \Big/ \sqrt{\tilde{k}(x,x)\tilde{k}(x',x')}$ and used $k(x, x')$ as a kernel function in our ML procedures.

5.1.2 Used device and qubits for the experiment

For the Fig. 2 of main text, we utilized the *ibm_sherbrooke* device and used the 12 qubits corresponding to [87, 88, 89, 74, 70, 69, 68, 67, 66, 73, 85, 84].

5.1.3 Gate decompositions for Givens rotation.

In the experiment, a total of 36 Givens rotation blocks were needed, and each block required 2 CNOT gates (Supplementary Figure S7). As mentioned, for implementing on the IBM quantum hardware, we decompose CNOT and $R_Y$ into $\{R_Z(\theta), \mathrm{SX}, \mathrm{X}, \mathrm{ECR}(\pi/2)\}$

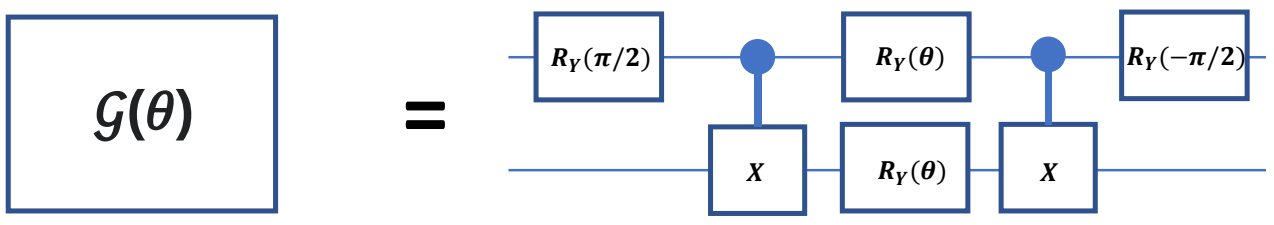

**Supplementary Figure S7 |** Givens rotation block ($\mathcal{G}(\theta) = \mathcal{U}(G(\theta))$) decompositions. We decomposed the Givens rotation block using two CNOTs. In the experiment, we implemented each CNOT by decomposing it into one ECR(π/2) and single qubit gates.

### 5.1.4 The process of selecting hyperparameter (λ) and the ML model performance from different hyperparameters

In the ML, we used a total of 200 data points for training ML model and 10,000 test data points for evaluating the model's performance. When determining hyperparameters λ in (2) of the main text, we used 1,000 validation data points that did not overlap with the test data. Among [0.0125, 0.025, 0.05, 0.125, 0.25, 0.5, 1, 2, 4, 8], we chose the λ that produced the smallest Root-Mean-Square Error (RMSE) on the validation data. We also trained ML model with non-optimal hyperparameter λ and observed that training generally worked well across a wide range of λ (Supplementary Figure S8). We used the hyperparameter λ = 0.05 in the Fig. 2 of the main text.

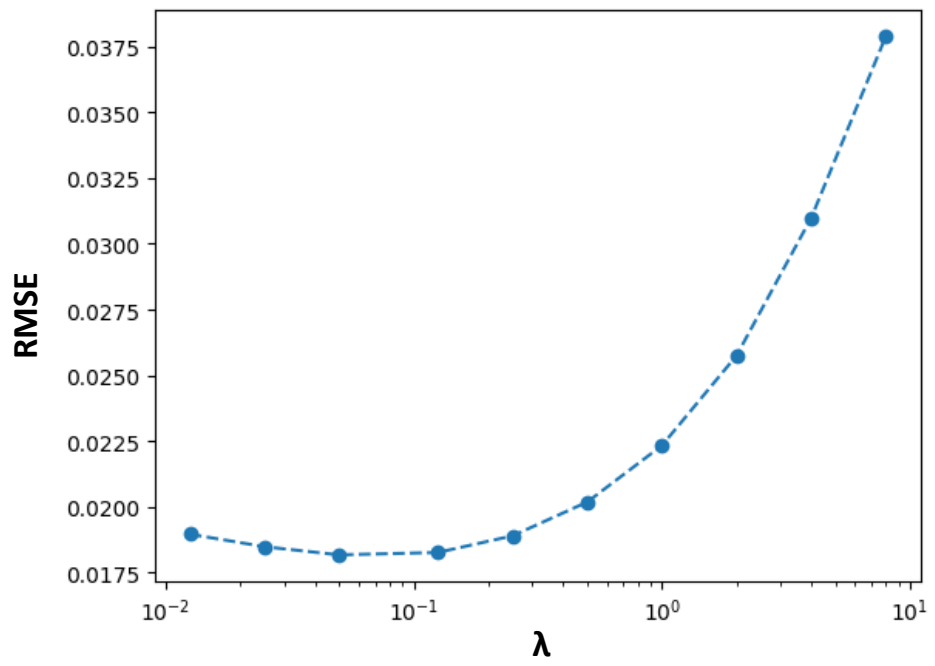

**Supplementary Figure S8** | Monitoring ML performance as we varied the hyperparameter (λ).

### 5.1.5 Edge correlation of Su–Schrieffer–Heeger (SSH) system.

The SSH Hamiltonian $H_{\text{SSH}} = \sum_i (va_{2i-1}^{\dagger}a_{2i} + wa_{2i}^{\dagger}a_{2i+1}) + h.c.$ exhibits either trivial or topological phases depending on the relative magnitudes of $v$ and $w$ in the Hamiltonian. Among the order parameters to distinguish these phases, edge correlation $a_1^{\dagger}a_n$ was employed in the Fig. 2g. By fixing $v = 1$ and varying $w$ between 0.25 and 1.75, we calculated the value of $\langle a_1^{\dagger}a_n \rangle$ with $n =$ 100 sites and $n = 12$ sites (used in the experiment of the main text) by classical simulation. As shown in the Supplementary Figure S9, there is a sharp change near the $w = 1$ which is consistent with known phase boundary between trivial and topological phase (trivial when $v < w$ and topological when $v > w$).

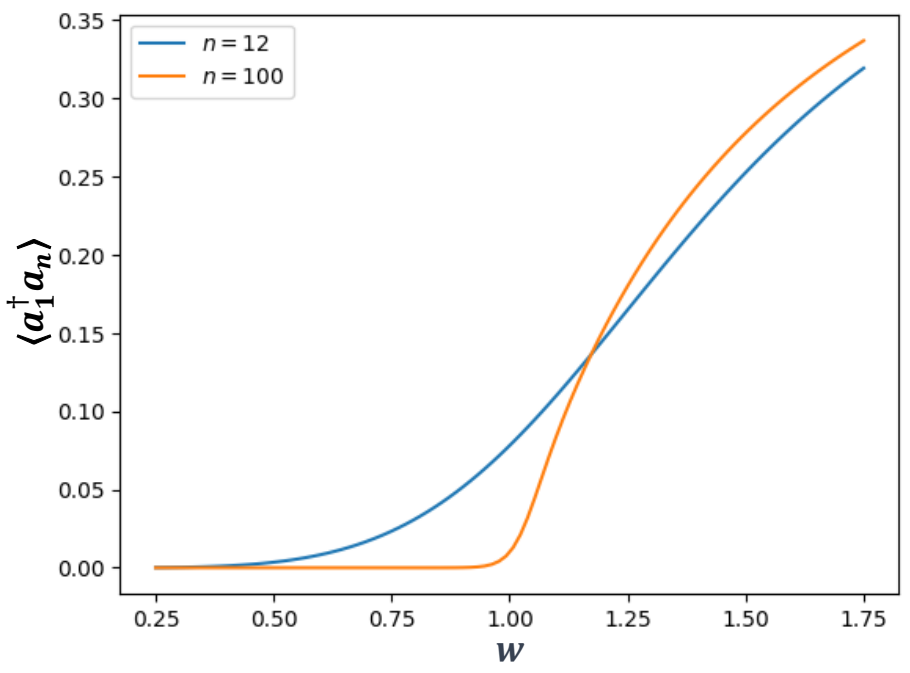

**Supplementary Figure S9** | Edge correlation $\langle a_1^{\dagger}a_n \rangle$.

5.1.6. Noisy simulation results

We conducted noisy simulations for the experiments in the main text. We investigated how the performance of the ML model changed as gate and measurement errors increase. In noisy simulation, in addition to the existing hardware errors, we added depolarizing error channels with error rate of $p_{\text{single}}$ and $p_{\text{two}}$ to single-qubit and two-qubit gates and measurement error with error rate $p_{\text{m}}$. Through this simulation, we can confirm that using various error mitigation techniques introduced in Supplementary Note 3 leads to better ML model performance (Supplementary Figure S10).

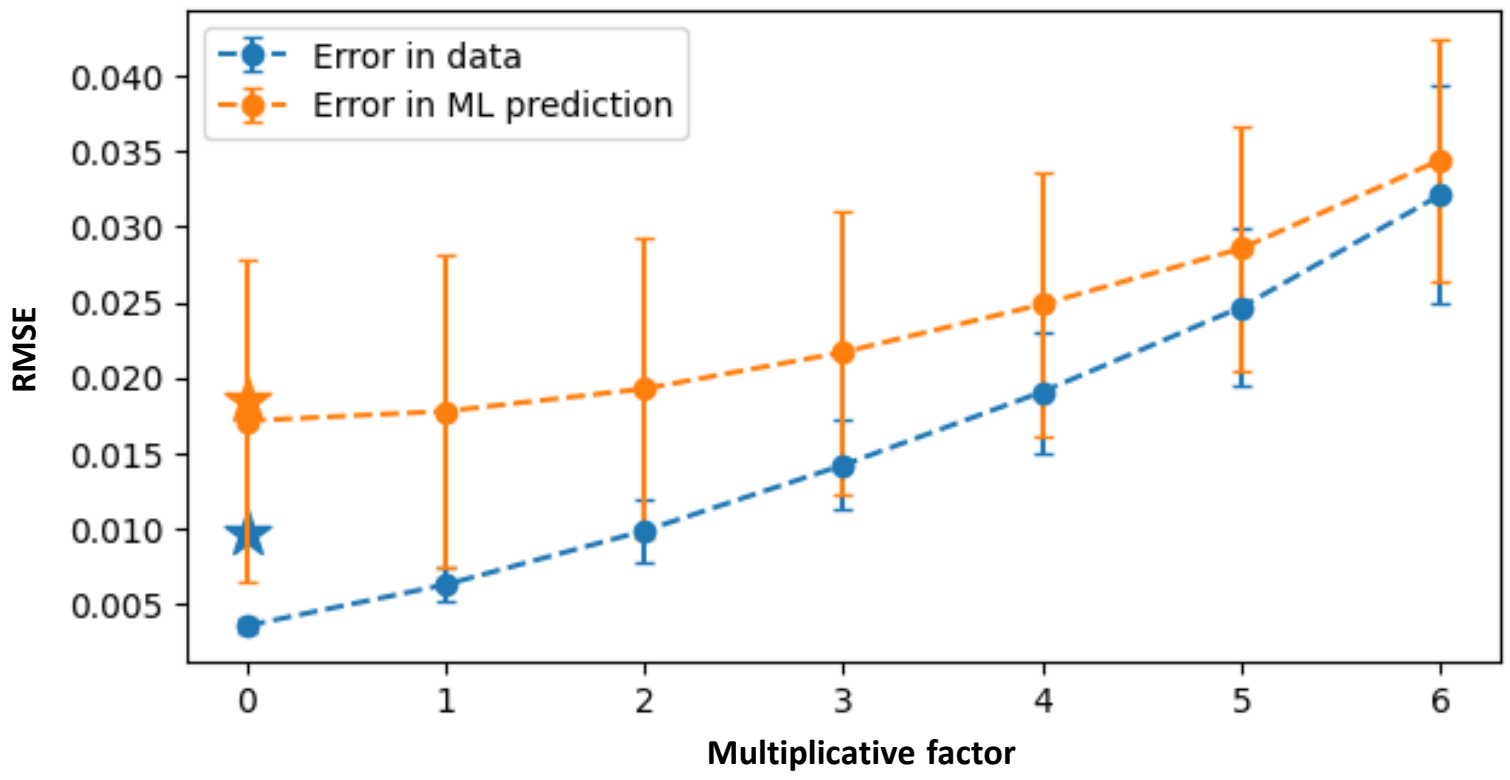

**Supplementary Figure S10** | Noisy simulation. The blue dots represent the RMSE for training data, while the orange dots indicate the RMSE for the 10,000 test data points obtained from the trained ML model. The asterisks represent the RMSE in the training data and ML prediction results used in the main text. The error bars represent the standard deviation of 100 instances. In the noisy simulation, we imposed ($p_{\mathrm{single}}$, $p_{two}$, $p_{\mathrm{m}}$) = (0.001, 0.01, 0.01) × (multiplicative factor).

## *Classifying quantum phase*

In the main text, we first conducted experiments to distinguish between SPT and trivial phases, and then, topologically ordered and trivial phases. Here, we will describe the details of the experiments and ML processes. For both experiments, we used kernel principal component analysis (PCA) and support vector machine (SVM) as classical ML algorithms. These two methods have been widely used in data analysis[19]. We first performed kernel PCA using the shadow kernel[10] developed for learning non-linear quantities of the quantum state $\rho$. After performing kernel PCA, each data point can be represented using principal eigenvectors as the bases. We focused up to the second principal components, reducing the input data dimension to 2. In the 2D space, we executed kernel SVM with the Gaussian kernel, utilizing the software package sklearn[30]. Through ML, we can obtain the phase boundary in the 2D projected space, and this enables us to predict the phase of a new quantum state.

5.2. Distinguishing a short-range entangled state from a trivial one

5.2.1. Used device and qubits for the experiment

For the images in Fig. 3 of the main text, we utilized the *ibm_sherbrooke* device and used the 44 qubits corresponding to [126, 112, 108, 107, 106, 105, 104, 103, 102, 101, 100, 99, 98, 91, 79, 80, 81, 72, 62, 61, 60, 59, 58, 71, 77, 76, 75, 90, 94, 95, 96, 109, 114, 115, 116, 117, 118, 119, 120, 121, 122, 123, 124, 125].

5.2.2. Expectation values of stabilizers of the resulting cluster state.

We used the 1D cluster state corresponding to the ground state of $H_{ZXZ} = -\sum_i Z_{i-1}X_iZ_{i+1}$ with periodic boundary condition as the fixed points of the SPT phase. To verify how well the state was prepared on a quantum computer, we conducted an experiment measuring stabilizers $\langle Z_{i-1}X_iZ_{i+1}\rangle$ shown in the Supplementary Figure S11.

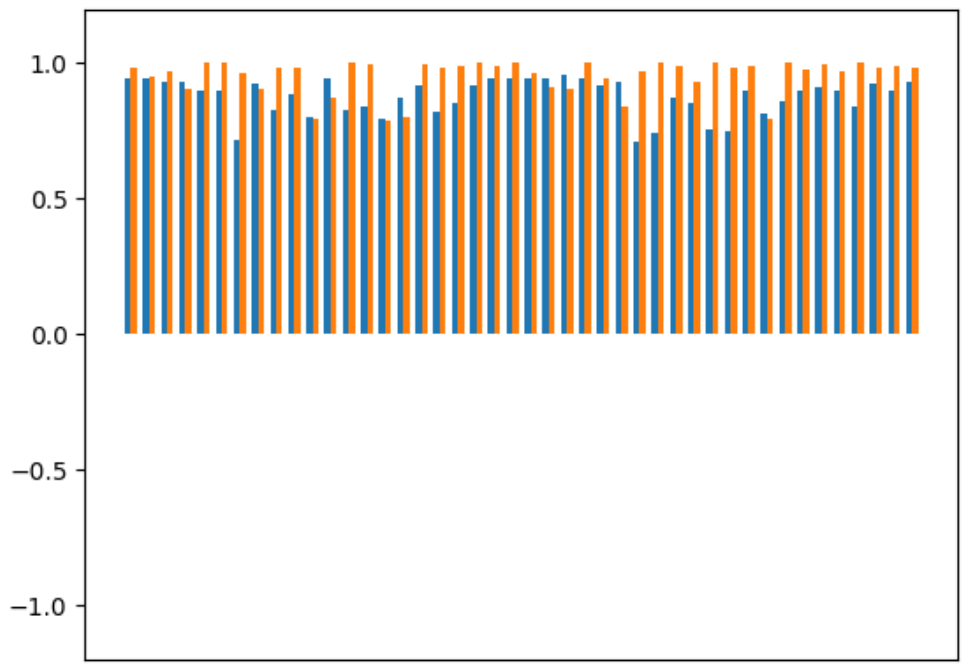

**Supplementary Figure S11** | Measurement result of the stabilizers for cluster state. The blue bar represents the results obtained from raw data, and the orange bar corresponds to the measurement-error-mitigated results. Through measurement error mitigation, we can compensate for the errors occurred during the measurement, as the average value shifts from 0.876 to 0.953.

5.2.3. Random unitary for generating data

In the experiment of main text, we used a total of 20 data points for training ML model, with 10 points each from the SPT and trivial phases. We employed the SPT phase protected by $\mathbb{Z}_2\otimes\mathbb{Z}_2$ symmetry generated by $X_{\text{even(odd)}} = \prod_{i=\text{even(odd)}} X_i$ or time reversal symmetry (TRS) generated by $\mathcal{T} = (\prod_i X_i)K$ where $X_i$, $Z_i$ are Pauli operators at the site $i$ and $K$ denotes complex-conjugation. We created different states within the same phase by applying single or two qubit gates sampled from certain set that respects the given symmetry[9]. In the case of $\mathbb{Z}_2\otimes\mathbb{Z}_2$ symmetry, we uniformly sampled $P\in\{I, X_1, X_2, X_1X_2\}$ and $\theta\in[0, 2\pi]$, and then applied $e^{i\theta P}$ to the corresponding qubits. For the TRS case, we followed the same approach, but sampling $P \in \{I, Z_1, Z_2, Z_1Y_2, Z_2Y_1, Z_1X_2, X_1Z_2\}$. We applied two layers of random unitary in each symmetry case. In particular, for the TRS case, it has been proven that when random unitary is sampled from the aforementioned set, it is hard to distinguish between the SPT and trivial phases using quantities in the form of $\mathrm{Tr}(O\rho)$[9]. Moreover, in actual experiments, the SOP value exponentially decreases as the length of SOP increases due to the measurement and gate errors, making distribution of SOP even denser near 0.

5.2.4. Hyperparameter in ML.

The shadow kernel in (SE7) has two hyperparameters, $\tau$ and $\gamma$, and in the main experiment, we used $\tau = \gamma = 1$. After kernel PCA, kernel SVM with a Gaussian kernel was employed with hyperparameter $\alpha = 1/(n_{\text{feature}}\sum_{ij}|\mathrm{X}_{ij}\text{-}\mathrm{E}(\mathrm{X})|^2)$, where $n_{\text{feature}} = 2$ in our case, $\mathrm{E}(\mathrm{X}) = 1/(n_{\text{feature}}\cdot N_{\text{data}})\sum_{ij}\mathrm{X}_{ij}$, and $\mathrm{X}_{ij}$ is the $j$-th feature value of the $i$-th data. Additionally, we investigated how the the phase boundary changed as we varied the hyperparameter $\alpha$ in the

Supplementary Figure S12.

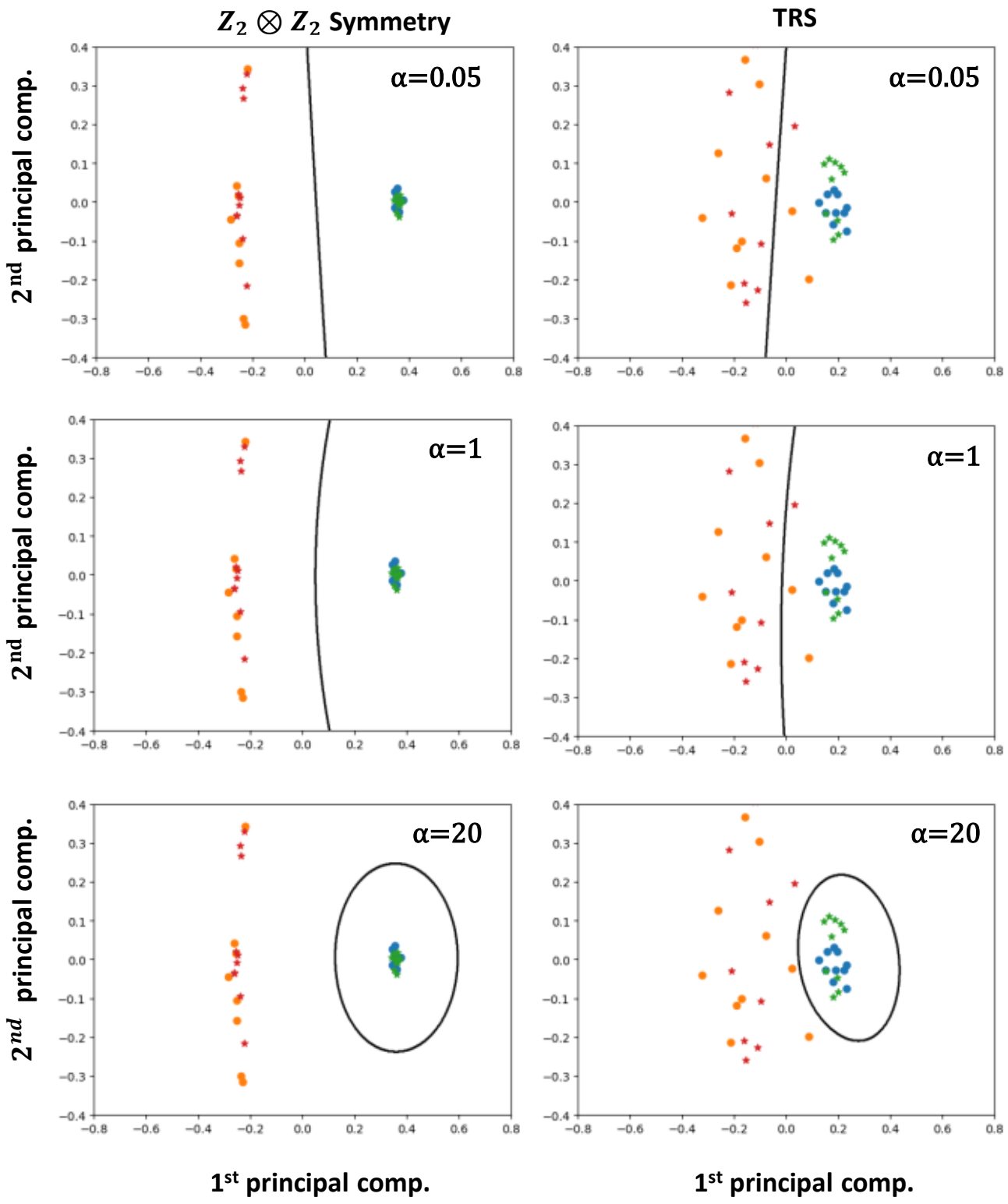

**Supplementary Figure S12** | Change of phase boundary depending on the hyperparameter α of the Gaussian kernel.

For Fig. 3e and 3f of the main text, the hyperparameter of Gaussian kernel is α = 7.87 in the case of $\mathbb{Z}_2 \otimes \mathbb{Z}_2$ symmetry and α = 18.8 in the TRS case.

## 5.2.5. Noisy simulation

We conducted noisy simulations for the experiments in the main text. To carry out noisy simulations, we employed the depolarizing error channel $D_p(\rho) = (1\text{-}p)\rho + pI_d/d$ where $d = 2^n$ and $I_d$ is $d \times d$ identity matrix. We investigated how the the distribution of the data changed as we varied depolarizing error rate $p$ (Supplementary Figure S13).

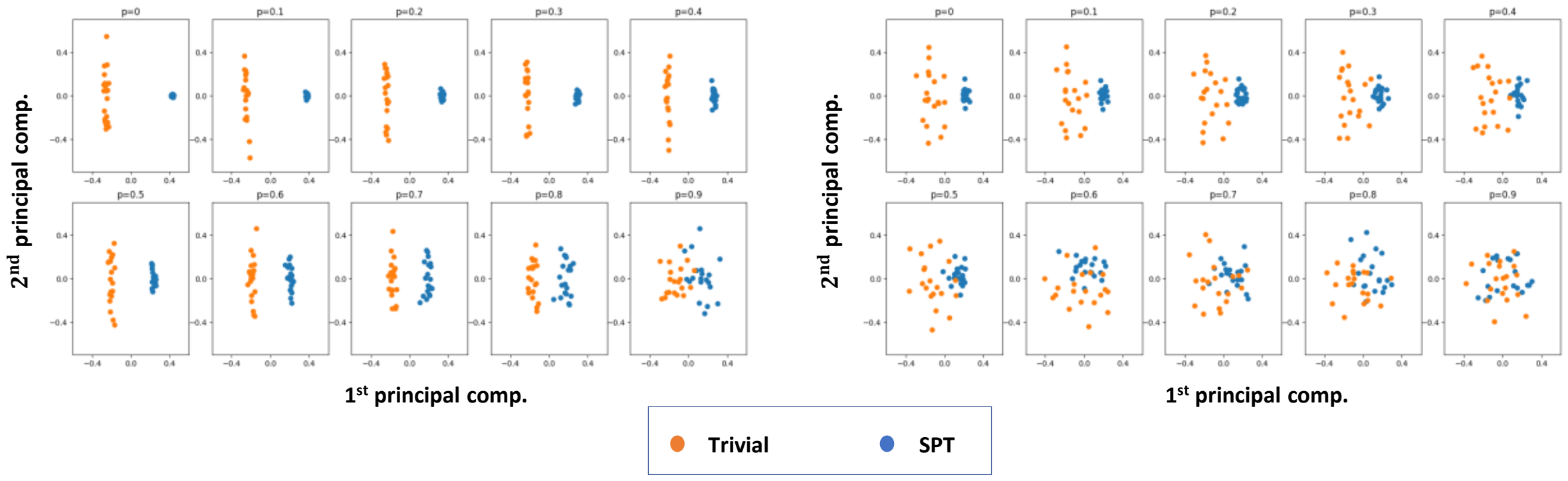

**Supplementary Figure S13** | **a**, Change of data distribution for $\mathbb{Z}_2 \otimes \mathbb{Z}_2$ symmetry case depending on the depolarizing error rate $p$. **b**, Change of data distribution for time reversal symmetry (TRS) case depending on the depolarizing error rate $p$.

## 5.3. Distinguishing a topological phase from trivial phase

### 5.3.1. Used device and qubits for the experiment

In the experiment for distinguishing between topologically ordered and trivial phases, we used *ibm_sherbrooke* and employed 61 qubits corresponding to [77, 76, 101, 102, 103, 105, 106, 107, 78, 92, 93, 79, 80, 81, 82, 83, 84, 85, 86, 87, 88, 89, 72, 73, 74, 60, 61, 62, 63, 64, 65, 66, 67, 68, 69, 70, 53, 54, 55, 41, 42, 43, 44, 45, 46, 47, 48, 49, 50, 51, 34, 35, 36, 23, 24, 25, 29, 28, 27, 31, 32]. Among them, 25 qubits were utilized to store information for the $|0_L\rangle$ with a code distance of 5, while the remaining 36 qubits were used as ancilla qubits (Supplementary Figure S14).

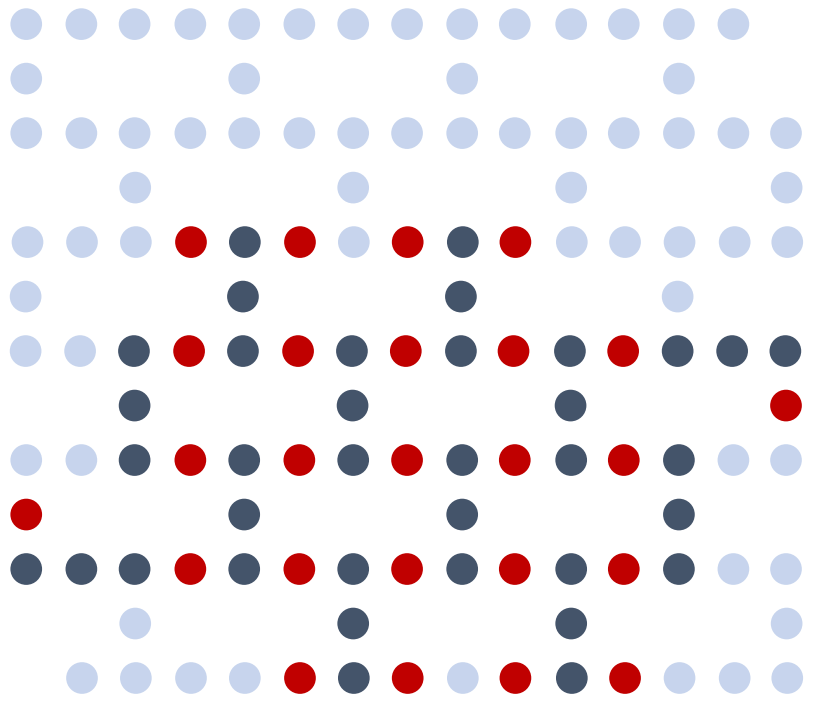

**Supplementary Figure S14** | Used 61 qubits for the experiment out of 127 qubits in *ibm_sherbrooke*. The 25 red

colored qubits were used to store information for the $|0_L\rangle$ state with $d_{code} = 5$ while the remaining 36 qubits served as ancilla qubits needed for measurement-assisted state preparation method.

### 5.3.2. Stabilizer measurement result

We used the $|0_L\rangle$ state in (4) of the main text corresponding to one of the ground states of $H = -\sum_s A_s - \sum_p B_p$, $(A_s = \prod_{i\in s} Z_i,\ B_p = \prod_{j\in p} X_j)$ where $s$ is a Z-plaquette and $p$ is a X-plaquette as the fixed-point state of the topologically ordered phase. To verify how well the state was prepared on a quantum computer, we conducted an experiment measuring stabilizers (Supplementary Figure S15).

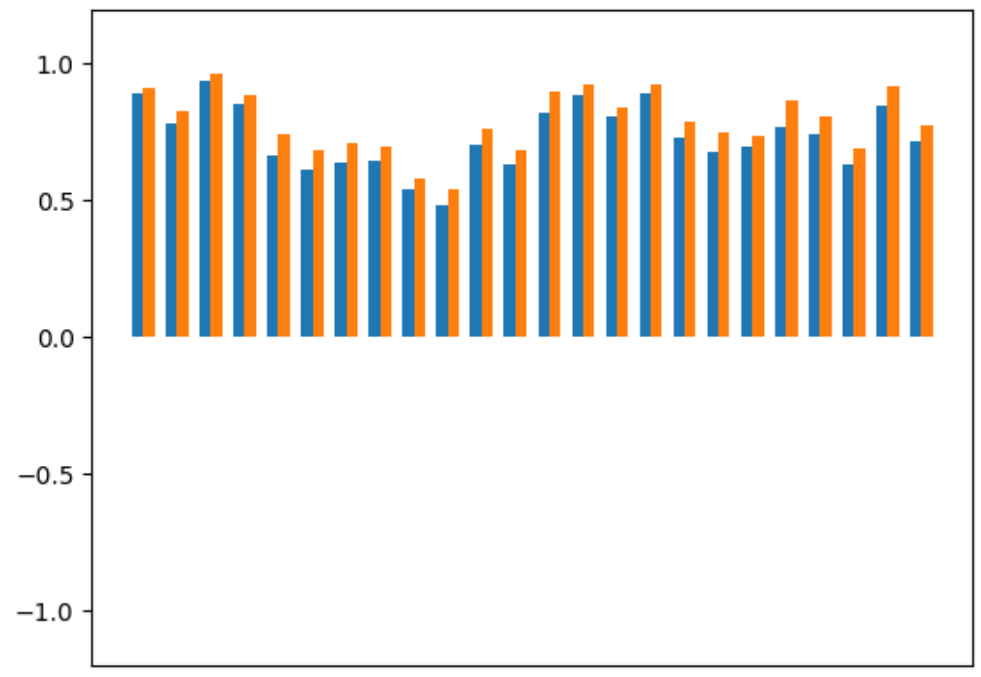

**Supplementary Figure S15** | Measurement result of the stabilizers for $|0_L\rangle$ state. The blue bars represent the results obtained from raw data, and the orange bars correspond to the measurement-error-mitigated results. Through measurement error mitigation, we can compensate for the errors occurred during the measurement, as the average value shifts from 0.733 to 0.784.

### 5.3.3. Random unitary for generating data

In the experiment of the main text, we applied local random unitary to create 10 different training data in each phase. In the topologically ordered phase, we applied the tensor product of single qubit gates due to hardware qubit connectivity, while in the trivial phase, we applied local random unitary up to 5 layers where each layer consists of CX gates and random single qubit gates. Although we only applied relatively simple tensor product of single qubit gates for generating data in the topologically ordered phase, it is known that it is impossible to distinguish a topologically ordered phase from a trivial phase using (global) linear observable corresponding to $\mathrm{Tr}(O\rho)$[10]. Therefore, using conventional topological string order parameters is highly likely to fail in cases like ours, where the target state deviates from the fixed-point state.

5.3.4. Hyperparameter in ML.

The shadow kernel in (SE7) has two hyperparameters, $\tau$ and $\gamma$, and in the experiment of main text, we used $\tau = \gamma = 1$.

5.3.5 Noisy simulation

We conducted noisy simulations for the experiments obtained from the quantum computer in the main text. To carry out noisy simulations, we employed the depolarizing error channel $D_p(\rho) = (1\text{-}p)\rho + pI_n/d$ where $d = 2^n$ and $I_n$ is $d \times d$ identity matrix. We investigated how the distribution of the data changed as we varied depolarizing error rate $p$ (Supplementary Figure S16).

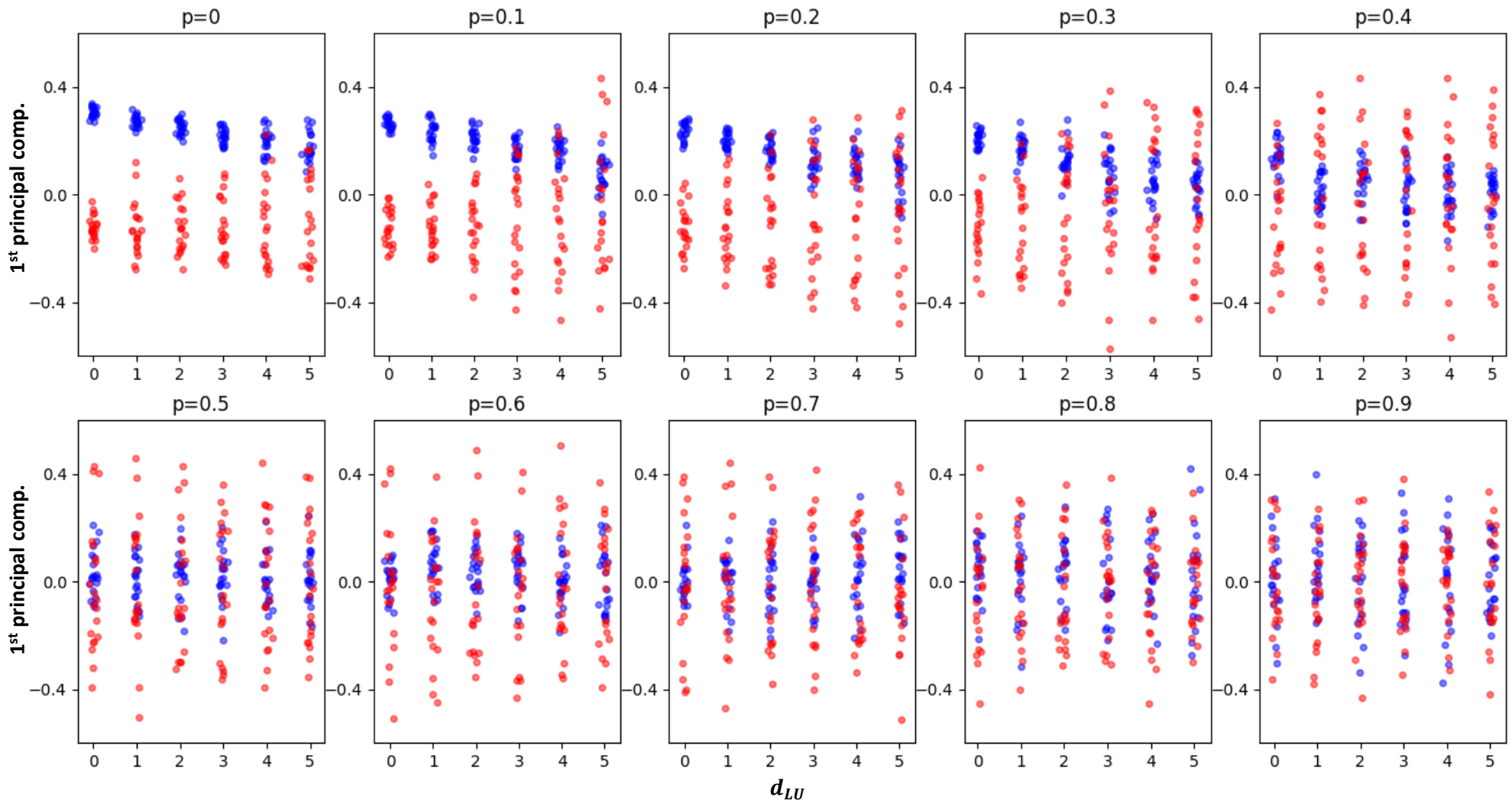

**Supplementary Figure S16** | Change of data distribution depending on the depolarizing error rate *p*.

## 5.4. Unlocking the black box of the ML model

Although our analysis is based on methodologies with mathematically guaranteed performance, there are instances in actual experiments or simulations where certain conditions deviate slightly from the assumptions used in mathematical proofs[10]. For instance, in the problem of predicting the properties of the ground state, the data used for training must all be in the same phase, and the system we want to predict through the ML model must also be in the same phase as training data to guarantee the proven mathematical performance. However, our ML task used data from a 1D nearest neighbor random hopping system containing multiple quantum phases, not strictly satisfying the required mathematical assumptions. Additionally, in the classification task, we used a modified shadow kernel (SE7) for numerical stability, which deviates slightly from the original one (3) in the main text. In such cases, we might question the performance of the ML model and may require information about what aspects of the data ML model has considered during the ML procedure. In this context, we conducted an experiment to extract what structure the ML model employed within the data in classifying between topologically ordered and trivial phases in a 9-qubit system.

5.4.1. Used device and qubits for the experiment

For the Fig. 5 of main text, we utilized the *ibm_sherbrooke* device and used the 9 qubits corresponding to [117, 118, 119, 120, 121, 122, 123, 124, 125] (Supplementary Figure S17).

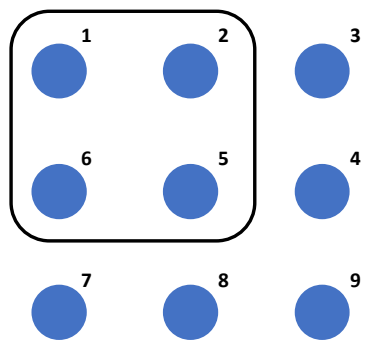

**Supplementary Figure S17** | Out of 9 qubits, the [1, 2, 5, 6] sub-system was used for ML

5.4.2. Random unitary for generating data

In the experiment, to create states in the topologically ordered and trivial phases, we applied a random unitary consisting of two layers to the fixed-point states of each phase. Each layer is composed of CX gates and arbitrary single qubit gates.

5.4.3. Feature mapping

To extract phase classifier from the ML model, we used the following feature vector $\varphi(\rho)\in R^{15}$, which maps the given state to a relatively lower-dimensional space.

$$\begin{aligned}\varphi(\rho) = [&S^{(2)}(\rho_{[1]}), S^{(2)}(\rho_{[2]}), S^{(2)}(\rho_{[5]}), S^{(2)}(\rho_{[6]}), S^{(2)}(\rho_{[1,2]}),\\ &S^{(2)}(\rho_{[1,5]}), S^{(2)}(\rho_{[1,6]}), S^{(2)}(\rho_{[2,5]}), S^{(2)}(\rho_{[2,6]}), S^{(2)}(\rho_{[5,6]}),\\ &S^{(2)}(\rho_{[1,2,5]}), S^{(2)}(\rho_{[1,2,6]}), S^{(2)}(\rho_{[1,5,6]}), S^{(2)}(\rho_{[2,5,6]}), S^{(2)}(\rho_{[1,2,5,6]})]^T\end{aligned} \quad \text{(SE20)}$$

where $\rho_A$ (= $\mathrm{Tr}_{\neg A}(\rho)$) is the reduced density matrix (RDM) obtained by tracing out all qubit indices except those corresponding to $A$ and $S^{(2)}(\rho) = -\log_2(\mathrm{Tr}(\rho^2))$ is a Renyi-2 entanglement entropy. In the experiment described in the main text, due to the relatively small number of qubits, the Maximum-likelihood estimator quantum state tomography (MLE-QST)[31] was used to obtain each RDM in the feature vector $\varphi(\rho)$.

5.4.4. Improvement in ML performance with measurement error mitigation (MEM).

We employed the MEM introduced earlier to offset the errors that occurred during the data acquisition for MLE-QST. By comparing the ML model's performance on test data after training with raw data and measurement-error-mitigated data, we confirmed that the successful phase classification rate increased from 0.942 to 0.959 when MEM was applied.

5.4.5. Comparison between the classifier obtained from ML and the Renyi-2 TEE.

To compare the performance of the classifier obtained from ML, $f_{\mathrm{ML}}(\rho) = w_{\mathrm{ML}}^{T}\varphi(\rho) + w_{0,\mathrm{ML}}$ ($w_{\mathrm{ML}}$ = [0.0780, 0.0736, 0.0232, 0.0342, 0.230, 0.103, 0.113, 0.0786, 0.0977, 0.091, 0.235, 0.254, 0.172, 0.0152, 0.171]$^{\mathrm{T}}$, $w_{0,\mathrm{ML}}$ = -2.23), and the existing $f_{\mathrm{TEE}}(\rho) = w_{\mathrm{TEE}}^{T}\varphi(\rho) + w_{0,\mathrm{TEE}}$ ($w_{\mathrm{TEE}} = [1,0,0,1,0,0,-1,1,0,0,-1,0,0,-1,1]^{T}$, $w_{0,\mathrm{TEE}} = 0.1$ which is set to minimize classification error), We generated 6000—3000 from each phase—test data via classical simulation. An error $\varepsilon \in [-0.1, 0.1]$ was added to each $S^{(2)}(\rho_A)$ of $\varphi(\rho)$ in (SE20) to emulate real-world experimental conditions.

**Supplementary References**